%% file: main.tex
\documentclass[epj,captions=tableabove]{svjour}
\usepackage{lipsum}%
\usepackage{multicol}%
\usepackage{graphics}
\usepackage{rotating}
\usepackage{amsmath}
\usepackage{subcaption}
\usepackage{upgreek}
\usepackage{xcolor}
\usepackage{comment}
\usepackage{textcomp}
\usepackage{gensymb}
\usepackage[normalem]{ulem}
\usepackage{siunitx}
\usepackage{mathtools,amssymb,nccmath}
\usepackage{threeparttable}

\usepackage{cuted}

\begin{document}

\title{STRASSE: A Silicon Tracker for Quasi-free Scattering Measurements at the RIBF}

\author{H.~N.~Liu\inst{1,2}\and F.~Flavigny\inst{3} \and H.~Baba\inst{4} \and M.~Boehmer\inst{5}\and U.~Bonnes\inst{1} \and V.~Borshchov\inst{6}\and P.~Doornenbal\inst{4} \and N.~Ebina\inst{7}\and M.~Enciu\inst{1} \and A.~Frotscher\inst{1}\and R.~Gernhäuser\inst{5}\and V.~Girard-Alcindor\inst{1}\and D.~Goupillière\inst{3}\and J.~Heuser\inst{8}\and R.~Kapell\inst{8}\and Y.~Kondo\inst{7}\and H.~Lee\inst{7} \and J.~Lehnert\inst{8} \and T.~Matsui\inst{7} \and A.~Matta\inst{3}\and T.~Nakamura\inst{7} \and A.~Obertelli\inst{1} \and T.~Pohl\inst{1} \and M.~Protsenko\inst{6} \and M.~Sasano\inst{4}\and Y.~Satou\inst{7} \and C.~J.~Schmidt\inst{8}\and K.~Schünemann\inst{8}\and C.~Simons\inst{8} \and Y.~L.~Sun\inst{1} \and J.~Tanaka\inst{4}\and Y.~Togano\inst{9}\and T.~Tomai\inst{7}\and I.~Tymchuk\inst{6}\and T.~Uesaka\inst{4}\and R.~Visinka\inst{8} \and H.~Wang\inst{7} \and F.~Wienholtz\inst{1}}

\institute{Institut f\"ur Kernphysik, Technische Universit\"at Darmstadt, 64289 Darmstadt, Germany \and Key Laboratory of Beam Technology of Ministry of Education,
College of Nuclear Science and Technology, Beijing Normal University, Beijing 100875, China\and Normandie Univ, ENSICAEN, UNICAEN, CNRS/IN2P3, LPC Caen, 14000 Caen, France\and RIKEN Nishina Center, 2-1 Hirosawa, Wako, Saitama 351-0198, Japan \and Department of Physics, Technical University of Munich, D-85748 Garching, Germany \and LTU, Kharkiv, Ukraine\and Department of Physics, Tokyo Institute of Technology, 2-12-1 O-Okayama, Meguro, Tokyo, 152-8551, Japan \and GSI Helmoltzzentrum f\"ur Schwerionenforschung GmbH, 64291 Darmstadt, Germany \and Department of Physics, Rikkyo University, 3-34-1 Nishi-Ikebukuro, Toshima, Tokyo 172-8501, Japan}

\date{31 December 2022}
\abstract{
STRASSE (Silicon Tracker for RAdioactive nuclei Studies at SAMURAI Experiments) is a new detection system under construction for quasi-free scattering (QFS) measurements at 200-250 MeV/nucleon at the RIBF facility of the RIKEN Nishina Center. It consists of a charged-particle silicon tracker coupled with a dedicated thick liquid hydrogen target (up to 150-mm long) in a compact geometry to fit inside large scintillator or germanium arrays.
Its design was optimized for two types of studies using QFS: missing-mass measurements and in-flight prompt $\gamma$-ray spectroscopy.
This article describes (i) the resolution requirements needed to go beyond the sensitivity of existing systems for these two types of measurements, (ii) the conceptual design of the system using detailed simulations of the setup and (iii) its complete technical implementation and challenges. The final tracker aims at a sub-mm reaction vertex resolution and is expected to reach a missing-mass resolution below 2 MeV  in $\sigma$ for $(p,2p)$ reactions when combined with the CsI(Na) CATANA array.}

\maketitle

\section{Introduction}
\input{introduction.tex}

\section{STRASSE concept and simulations}
\input{Simulations.tex}

\section{STRASSE technical design}
\input{STRASSE_tracker.tex}

\section{Liquid hydrogen target}
\input{LH2_target.tex}

\section{Summary}
\input{Conclusions}

\section{Acknowledgments}
H. N. L., M. E., A. F., V. G.-A., A. O., T. P. and Y. L. S. acknowledge the support from the Deutsche Forschungsgemeinschaft (DFG, German Research Foundation) -- Project No. 279384907 -- SFB 1245. H. N. L. is supported by the Fundamental Research Funds for the Central Universities of China. F. F. acknowledges the support from the Région Normandie (RIN Recherche 2019, Chaire d'excellence SIREN).

\end{document}

%% file: introduction.tex
\label{sec_intro}

Exploring nuclear shell evolution towards the neutron drip-line is a major focus in modern nuclear physics, since it provides benchmarks for our knowledge of the many-body forces at play inside the nucleus. With more accurate interactions, the structure of nuclei lying far away from the beta stability line as well as the location of the drip line could be better predicted. Such gains in predictive power are also essential to further understand and describe nucleosynthesis processes in the cosmos~\cite{erler12}. 

In the shell-model framework, the nuclear structure is determined by a delicate interplay between the spherical mean field (the monopole component of the Hamiltonian) and the multipole correlations.
In the conventional image, nuclei with magic numbers (2, 8, 20, 28, 50, 82) of protons or neutrons are spherical and associated with a large shell gap around the Fermi surface provided by the monopole Hamiltonian, while showing reduced correlations. However, progress in the past decades shows that the nuclear shell structure changes sometimes dramatically with proton-to-neutron asymmetry, and quadrupole correlations can become dominant even in presumably closed-shell nuclei, leading to a deformed ground state \cite{Otsuka2020,Nowacki2021}. Magic numbers can thus be considered as local features, sometimes less robust than anticipated. A few prominent examples are the breakdown of the traditional \emph{N} = 8 \cite{navin00} \cite{Chen2018}, 20 \cite{thi75,gui84}, 28 \cite{bastin07} magic numbers, the emergence of new magic numbers at N=16 \cite{tshoo12}, 32 \cite{huck85,wie13,Steppenbeck2015}, 34 \cite{step13,Michimasa2018,liu2019,Chen2019}, and the competition between spherical and deformed configurations in neutron-rich N = 50 nuclei \cite{yang2016,taniuchi2019}. Extensive efforts shed light on the driving forces behind such shell evolution. In particular, the spin-isospin terms of the monopole part of the effective nucleon-nucleon interactions \cite{tal60,zuker1994,otsuka05}, as well as three-nucleon forces originating from the composite nature of nucleons \cite{zuker03,otsuka10} have been proposed to contribute strongly to the modification of magic numbers. \\

\begin{figure}[t!]
\begin{center}
\includegraphics[width=0.48\textwidth]{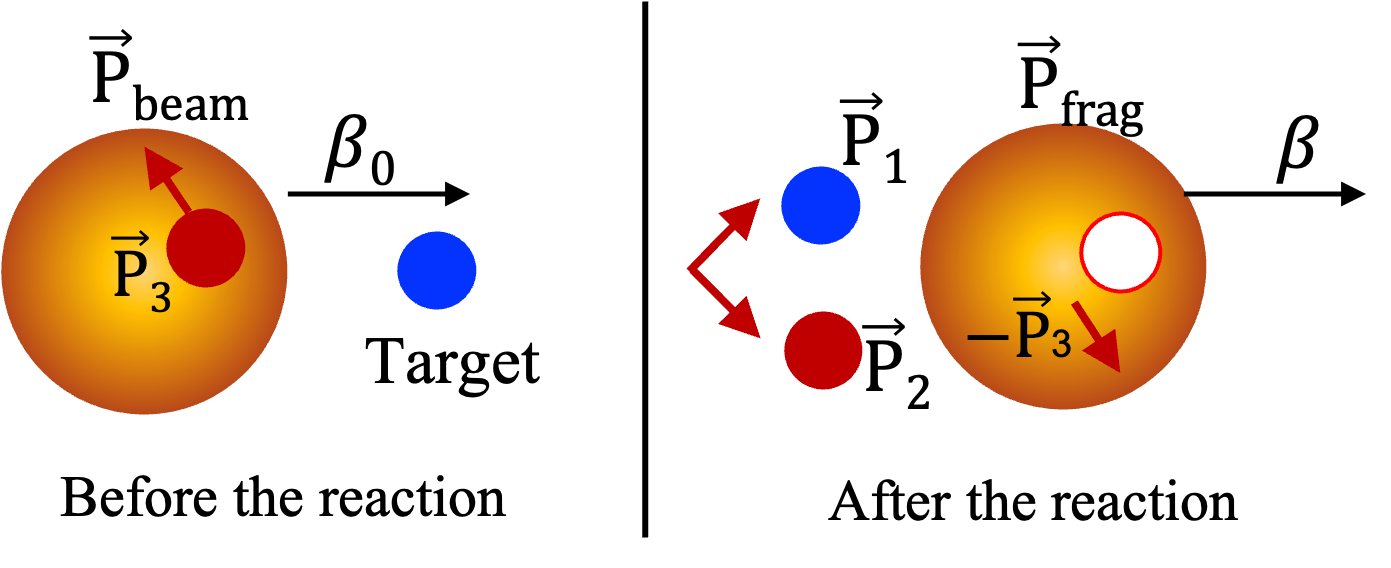}
\caption{\label{p2p} Schematic of the ($\emph{p}$,~2$\emph{p}$) quasi-free scattering reaction in inverse kinematics. $\vec{P}_\text{beam}$, $\vec{P}_\text{frag}$, $\vec{P}_{1}$, $\vec{P}_{2}$ are the momentum of the incident beam, outgoing fragment and two recoil protons in the laboratory frame. $\vec{P}_{3}$ is the intrinsic (missing-) momentum of the proton inside the projectile. $\beta_{0}$ and $\beta$ are the velocity of the incoming beam and the fragment, respectively}.
\end{center}
\end{figure}

Quasi-free scattering (QFS) in inverse kinematics with proton targets has been proven to be an effective method to study the microscopic structure of atomic nuclei, in which heavy reaction residues exhibit spectator-like properties\,\cite{Panin2016}. This technique has been applied to study shell evolution as well as single-particle structure, clustering, and short-range correlations in nuclei via in-beam gamma and/or missing-mass spectroscopy\,\cite{Panin2021}, revealing a variety of properties of nuclei with asymmetric neutron-to-proton ratio. However, improving the experimental sensitivity on the various observables obtainable from QFS measurements requires the development of highly granular, compact and weakly interceptive tracking systems to determine accurately the trajectories of the light reaction products (target-like protons and knocked-out fragments). In this article, we present the design of a new detection system called STRASSE (\underline{S}ilicon \underline{T}racker for \underline{RA}dioactive-nuclei \underline{S}tudied at \underline{S}AMURAI \underline{E}xperiments). The goal of this new device combining a thick liquid hydrogen (LH$_{2}$) target and a compact proton tracker is to allow detailed spectroscopy of very exotic nuclei produced at the RIBF via two methods: missing-mass measurements using ($\emph{p}$, 2$\emph{p}$) or ($\emph{p}$, 3$\emph{p}$) reactions (see Fig.\,\ref{p2p}) and high-resolution in-beam $\gamma$-ray spectroscopy. The conceptual requirements to go beyond the capabilities of existing setups are first detailed in this section. The concept and simulated performances of STRASSE are discussed in Section 2. The technical details of the tracker, including its mechanical design and electronics, are given in Section 3. The LH$_2$ target is  introduced in Section 4.\\

\subsection{Missing-mass spectroscopy requirements}
\label{sec:mmreq}

\subsubsection{Missing mass}
One of the main asset of QFS studies is the possibility to determine the excitation energy $E_x$ of the reaction residue. This method is called \emph{missing mass} and has been applied for example to search for the tetraneutron~\cite{Duer2022} and to study $\alpha$-clustering at the surface of Sn isotope \cite{Tanaka2021}. In ($\emph{p}$, 2$\emph{p}$) reactions, the missing-mass measurement relies on the determination of the Lorentz momenta of two recoil protons and of the incident beam. $E_x$ is then calculated using the following formulae:

\begin{equation} \label{miss_mass}
\begin{split}
E_x & =  \sqrt{(E_\text{beam}+E_\text{tgt}-E_{1} -E_{2})^2-(\vec{P}_\text{beam}-\vec{P}_{1}-\vec{P}_{2})^2} \\
& -M_\text{frag},
\end{split}
\end{equation} 
where $E_\text{beam}$, $E_\text{tgt}$, $E_{1}$, $E_{2}$ are the total energy of the beam, target and two recoil protons, respectively.
$\vec{P}_\text{beam}$, $\vec{P}_{1}$, $\vec{P}_{2}$ are the momentum of the incident beam and two recoil protons in the laboratory frame.  

From Eq. \ref{miss_mass}, one can derive the missing-mass energy resolution ($\sigma_{E_x}$) as a function of uncertainties on the scattering angle ($\sigma_\theta$), the beam momentum ($\sigma_{P_\text{beam}}$) and the kinetic energy of the recoil protons ($\sigma_T$):
\begin{strip}
\begin{equation}
\begin{aligned}
\sigma_{E_x}^2 &= (p_{1}^2\beta_{0}^2\gamma^2\sin^2\theta_{1}+p_{2}^2\beta_{0}^2\gamma^2\sin^2\theta_{2})\sigma_\theta^2  \\
& + \left(\frac{p_\text{beam} [p_1\cos\theta_1+p_2\cos\theta_2-\beta_{0}(T_1+T_2+m_p)]}{M_{\text{frag}}+E_x}\right)^2 \left(\frac{\sigma_{p_\text{beam}}}{p_\text{beam}}\right)^2\\
& + \left(\frac{(T_1+T_2+m_p-\gamma M_\text{beam}+[p_2\cos(\theta_1+\theta_2)+p_\text{beam}\cos\theta_1-p_1]/\beta_1) T_1}{M_\text{frag}+E_{x}}\right)^2\left(\frac{\sigma_T}{T_1}\right)^2\\
& + \left(\frac{(T_1+T_2+m_p-\gamma M_\text{beam}+[p_1\cos(\theta_1+\theta_2)+p_ \text{beam}\cos\theta_2-p_2]/\beta_2)T_2}{M_\text{frag}+E_x}\right)^2\left(\frac{\sigma_T}{T_2}\right)^2,
\end{aligned}
\label{miss_reso_eq}
\end{equation}
\end{strip}
where $\beta_{0}, M_\text{beam}$ are the velocity and the mass of the incoming beam. $T_{1} (T_{2})$, $p_{1} (p_{2}$), $\beta_1 (\beta_2)$ are the kinetic energies, momentum and velocities of the recoil protons. $m_{p}$ is the static mass of the proton. $\theta_1 (\theta_2)$ are the polar angles between scattered protons and the beam axis. $M_\text{frag}+E_x$ is the missing mass of the fragment. For ($\emph{p}$, 2$\emph{p}$) reactions at 250 MeV/u, typical contributions to the missing mass resolution ($\sigma$), arising from the uncertainty in scattering angle, the beam momentum and total kinematic energy of protons are: 
\begin{itemize}
\item $\sigma_{E_x}/\sigma_\theta \sim$ 0.3 MeV/mrad; 
\item $\sigma_{E_x}/(\sigma_{p_\text{beam}}/p_\text{beam})\sim$ 0.2 MeV/\%;
\item $\sigma_{E_x}/(\sigma T/T)\sim$ 0.3 MeV/\%.
\end{itemize}
Experimentally, $\sigma_{p_\text{beam}/p_\text{beam}}$and $\sigma_T/T$ can be kept well below 1\% and are not the dominating factors. For example at the RIBF facility, the typical beam momentum resolution of the BigRIPS spectrometer~\cite{BigRIPS} extracted from previous measurements is $\sim$ 0.1\% ($\sigma$) and the energy resolution of CsI(Na) crystals from the CATANA array~\cite{catana,Nishimura2020} used to measure the energy of scattered protons is approximately 0.7\% ($\sigma$) for 100-MeV protons. As a consequence, the missing-mass resolution is dominated by the precision reachable on the angles of the two protons. 

Up to now, MINOS~\cite{obertelli14,Santamaria18} is the main apparatus used at the RIBF to populate states in rare isotopes after such proton-induced one or two nucleon removal. It is composed of a thick LH$_{2}$ target (up to 150-mm long) and a Time Projection Chamber (TPC) for proton tracking. In case of ($\emph{p}$,~2$\emph{p}$) reactions, two recoil protons in the QFS with large momentum transfer are emitted, centered around 45 degrees in the laboratory frame, as schematically shown in Fig.\,\ref{p2p}. The tracks of outgoing protons were recorded by the cylindrical TPC surrounding the target to reconstruct reaction vertices. However, the MINOS setup was mainly designed for in-beam $\gamma$-ray spectroscopy (see next section) and is not well suited for missing mass studies because of the large angular straggling of the recoil protons in the LH$_{2}$ target with a diameter of 52 mm and materials lying between the target and TPC (2mm thick Al chamber, 3 mm Rohacell and 125\,$\upmu$m Kapton). As shown in Table \ref{tab:straggling}, the total angular straggling of recoil protons with kinematic energy of 125 MeV emitted at 45 degrees passing by the MINOS setup is about 16.2 mrad in $\sigma_{\theta}$, leading to a missing mass resolution of 5 MeV in $\sigma$.

\begin{table}
\caption{Comparison of the angular straggling ($\sigma_{\theta}$) of recoil protons with kinematic energy of 125~MeV emitted at 45 degrees passing by the MINOS setup and a system including an LH$_{2}$ target and a silicon tracker in vacuum. The calculations were done via LISE++~\cite{Tarasov2016}.}

 \begin{threeparttable}
 \centering
\begin{subtable}[c]{0.5\textwidth}
\centering
\begin{tabular}{c|c}
    \hline
    \hline
    \multicolumn{2}{p{8.0 cm}} {\centering LH$_{2}$ target and TPC (MINOS)}\\
    \hline
        Factor & $\sigma_{\theta}$ (mrad) \\
    \hline
    LH$_{2}$ & \\
    \hline
    R = 26 mm & 4.0 \\    
    150 $\upmu$m Mylar foil & 1.6 \\   
    \hline
    TPC & \\
    \hline
    Al chamber (2.0 mm) & 10.8\\
    Rohacell (3.0 mm) & 1.7 \\
    Kapton ( 125 $\upmu$m) & 1.5\\
    Position resolution of TPC & 11.0\\  
    \hline
    \textbf{Total}  & \textbf{16.2}\\
    \hline
    \hline
    \end{tabular}
\end{subtable}

\vspace*{0.5 cm}

\begin{subtable}[c]{0.5\textwidth}
\centering
 \begin{tabular}{c|c}
    \hline
    \hline
    \multicolumn{2}{p{8.0cm}}{\centering LH$_{2}$ target and Si tracker}\\
    \hline
        Factor & $\sigma_{\theta}$ (mrad) \\
    \hline
    LH$_{2}$ & \\
    \hline
     R = 10 mm & 2.5 \\    
     \textbf{R = 15.5 mm} & \textbf{3.0} \\    
    \textbf{175 $\upmu$m Mylar foil} & \textbf{1.7} \\   
    \hline
    In-vacuum Si tracker & \\
    \hline
    Thickness of the inner layer Si & \\
    50 $\upmu$m & 1.6\\   
    100 $\upmu$m & 2.3\\
    \textbf{200 $\upmu$m} & \textbf{3.3}\\ 
    Pitch size of the inner layer Si & \\ 
     100 $\upmu$m & 0.5\\
     \textbf{200 $\upmu$m} & \textbf{1.0}\\
     300 $\upmu$m & 1.4\\
    \hline 
    \textbf{Total\tnote{a}} & \textbf{4.9}\\
    \hline
    \hline
    \end{tabular}
\begin{tablenotes}
\footnotesize
\item[a]{Total effect for the STRASSE setup including a LH$_{2}$ target with R = 15.5 mm and 200\,$\upmu$m thick inner layer Si with a pitch size of 200\,$\upmu$m.}
 \end{tablenotes}
 \end{subtable}
 \end{threeparttable}
  \label{tab:straggling}
\end{table}

To go beyond these missing-mass limitations with MINOS and reach a targeted $\sigma_{E_x}<2$\,MeV resolution, the new STRASSE system has to minimize the proton angular straggling in the materials between the reaction vertex and the tracker but also within the tracker. For example, 3.3\,mrad angular straggling is already induced in a 200\,$\upmu$m thick silicon as shown in Table \ref{tab:straggling} and leads to a resolution of 1\,MeV in $\sigma$. 
Therefore, the radius of the LH$_{2}$ target has to be as small as possible, the inner layer of the tracker has to be as thin as possible and the full STRASSE system has to be placed inside a vacuum chamber. In addition, in order to reach an effective design, the efficiency to reconstruct the reaction vertex and two proton tracks for ($\emph{p}$, ~2$\emph{p}$) reactions should be kept $\geq$ 80\% and $\geq$ 50\%, respectively. \\

\subsubsection{Missing momentum}

An important observable in the one-nucleon removal reaction is the momentum distribution of the struck nucleon whose shape is sensitive to its orbital angular momentum. In the projectile rest frame, the momentum of the removed nucleon is the same as that of the reaction fragment but in the opposite direction. For the ($\emph{p}$,~2$\emph{p}$) reaction in inverse kinematic, one can either measure the momentum of the reaction residue or reconstruct the missing momentum of the knocked-out proton using the momenta of two recoil protons (See Eq. \ref{mom_miss}). The perpendicular momentum of the removed nucleon can be directly extracted, while a Lorentz transformation has to be performed to obtain the longitudinal momentum distribution in the projectile rest frame with the following formulae:

\begin{equation}
\begin{split}
& P_{\perp}=(\vec{P}_\text{frag})_{\perp}
\\
&
P_{\parallel}=\frac{(\vec{P}_\text{frag})_{\parallel}-\beta_{0}E_\text{frag}}{1-\beta_{0}^2}
\\
&
P_{3{\perp}}=(\vec{P}_{1}+\vec{P}_{2})_{\perp}
\\
&
P_{3{\parallel}}=\sqrt{1-\beta_{0}^2}(\vec{P}_1+\vec{P}_2)_{\parallel}-\beta_{0}(M_\text{beam}-M_\text{frag}),
\label{mom_miss}
\end{split}
\end{equation} 
where $E_\text{frag}$, $\vec{P}_\text{frag}$, $M_\text{frag}$ are the total energy, momentum and mass of the fragment, respectively. $\beta_{0}$ and $M_\text{beam}$ are the velocity and mass of the projectile. $\vec{P}_{1}$ and $\vec{P}_{2}$ are the momentum of the two recoil protons as shown in Fig.\,\ref{p2p}.  
When using the momentum of the heavy fragment, the uncertainties on the beam and fragment velocities as well as on the angle between the fragment and the beam direction dominate the momentum resolution. In this case, the vertex tracking properties of the MINOS setup allows to achieve a decent momentum resolution (40~MeV/c in $\sigma$ for $^{52,54}$Ca($\emph{p}$,~2$\emph{p}$) reaction \cite{Sun2020}) even with a thick liquid hydrogen target. 

However, in many cases for exotic nuclei studies close or beyond the drip lines, the reaction fragments can be unbound or populated in unbound excited states. As a consequence, the fragment decays by emitting particles, and the exact reaction channel becomes harder to detect exclusively. In these cases, the use of the kinematics of the two scattered protons and the reconstructed missing momentum $P_3$ becomes a real asset. As shown by Eq.~\ref{mom_miss}, the missing momentum resolution is driven by the momentum determination of the projectile and two recoil protons. The missing momentum with the MINOS setup was not usable due to its poor angular resolution. STRASSE's targeted angular resolution of 5 mrad in $\sigma$ will overcome this limitation and enable missing-momentum studies with a resolution of $\sigma=$\,3\,MeV/c.\\

\subsection{In-beam $\gamma$-ray spectroscopy requirements}

In-beam $\gamma$-ray spectroscopy from knockout reactions of secondary beams produced by fragmentation or in-flight fission at intermediate energy often provides the first observable accessible by experiments to characterize the nuclear shell structure. Worldwide, four main laboratories (GANIL in France, GSI/FAIR in Germany, FRIB in the USA, and the RIBF of RIKEN in Japan) are currently using this technique. Notably, the RIBF facility of RIKEN \cite{Sakurai2008} and the future FAIR facility \cite{Spiller2006} will share the leadership in the next 10 years in producing radioactive beams at energies above 200\,MeV/u which will allow the use of thick secondary targets to balance the low beam intensities for the most exotic species. However, the high secondary beam energy also causes a large Doppler shift of the measured $\gamma$–ray energy relative to the de-excitation energy in the projectile rest-frame. The capability of a $\gamma$-ray detection setup to perform a good Doppler-shift correction is thus as important as its geometrical efficiency for the in-beam $\gamma$ measurement.

Doppler correction requires to measure i) the ejectile velocity ($\beta$) when emitting the transition and ii) the $\gamma$-ray emission angle ($\theta$) defined between the ejectile and the emitted $\gamma$-ray direction, as shown in the following equation:

\begin{equation}
E_0 = \frac{E_{\gamma} (1-\beta\cos{\theta})}{\sqrt{(1-\beta^2)}},
\label{in-beam-gamma}
\end{equation} 
where $\emph{E}_0$ is the nuclear transition energy in the rest frame of the ejectile, and $\emph{E}_\gamma$ is the measured $\gamma$-ray energy in the laboratory frame. 
On one hand, the uncertainty on $\beta$ essentially depends on the velocity spread inside the target due to energy-loss processes and, indirectly, on how precise the reaction vertex can be determined to perform such energy-loss corrections. On the other hand, the uncertainties on the angle $\theta$ are often dominated by the granularity of the $\gamma$-ray detector and the reaction vertex resolution. Therefore, for a given beam intensity, all existing setups are subject to subtle trade-offs between resolution and efficiency or luminosity: i) improving the angular/position resolution often comes at the cost of reducing efficiency for the $\gamma$-ray spectrometer and ii) thickening the target increases luminosity but degrades the Doppler correction due to energy losses and velocity spread within the target.

The energy resolution reachable after Doppler correction is directly related to Eq.~\ref{gamma_reso} and can be expressed as:

\begin{equation}\label{gamma_reso}
\begin{split}
\left(\frac{\Delta{E_0}}{E_0}\right)^2 
& = \left(\frac{\Delta E_{\gamma}}{E_{\gamma}}\right)^2+\left(\frac{\beta\sin\theta}{1-\beta\cos\theta}\right)^2\times\left(\Delta\theta\right)^2+
\\
&  \left(\frac{\beta-\cos\theta}{\left(1-\beta^2\right)\left(1-\beta\cos\theta\right)}\right)^2\times\left(\Delta\beta\right)^2,
\end{split}
\end{equation}
where $\Delta E_{\gamma}$ is the uncertainty on the measured $\gamma$ energy in the lab frame, $\Delta\theta$ is on the $\gamma$-ray emission angle and $\Delta\beta$ is on the erectile velocity.

\begin{figure*}[t!]
\begin{center}
\includegraphics[width=0.98\textwidth]{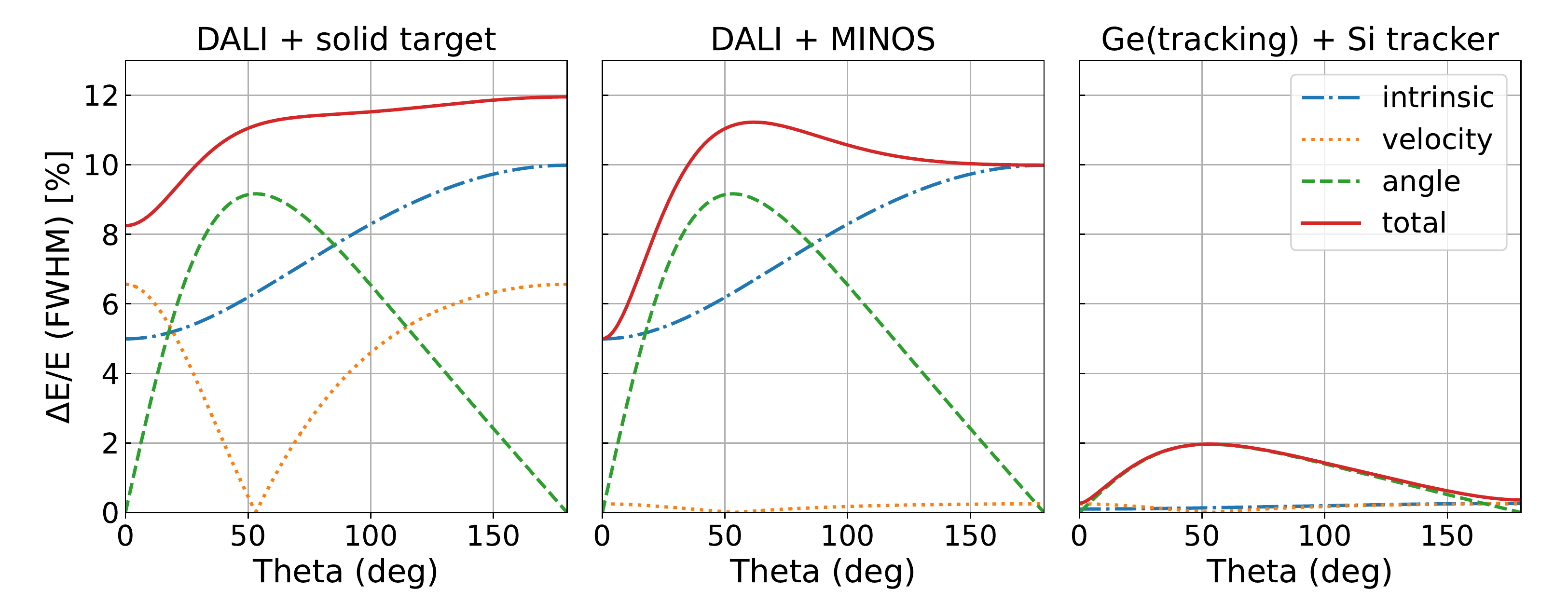}
\caption{\label{gamma_resolution} Comparison of Doppler-corrected energy resolution as a function of the $\gamma$-ray emission angle 
$\theta_{\gamma}$ for different experimental setups (solid lines). Individual contributions to the total from different sources of uncertainty are also displayed: angle $\Delta\theta$ (dashed lines), velocity $\Delta\beta$ (dotted lines), and intrinsic energy resolution $\Delta E_{\gamma}$ (dash-dotted lines). Panel (a) corresponds to a standard heavy ion target and DALI2 setup, (b) to the MINOS (LH$_{2}$+TPC) and DALI2 setup, and (c) to a new generation $\gamma$-tracking array used with a precise vertex tracker (see text for details).}
\end{center}
\end{figure*}

Figure \ref{gamma_resolution}(a) shows the energy resolution for the typical DALI2+solid target setup used at the RIBF-RIKEN to perform the spectroscopy of very exotic nuclei using nucleon removal reactions. It combines the use of a thick solid target [C, Be, (CH$_{2}$)$_n$] and the very efficient array of NaI(Tl) scintillators,  DALI2~\cite{dali2}. While thick targets and high-efficiency array favor luminosity and allow first spectroscopy of exotic nuclei produced at only a few particles per second, the achievable resolution is rather moderate ($\sim10\%$ FWHM at 1 MeV) and is dominated by the large average angular opening of the DALI2 crystals (7$^\circ$ FWHM) and their limited intrinsic energy resolution ($9\%$ FWHM at 0.662 MeV for the $^{137}$Cs standard source). Typical velocity uncertainty ($\Delta\beta/\beta\simeq$\,7$\%$~\cite{dali2}) originates essentially from the unknown vertex location within the solid target (assumed here to be 2.54 g/cm$^{2}$ thick C) but only plays a significant role in the total resolution at angles close to the beam axis.

As shown in Fig.\,\ref{gamma_resolution}(b), some limitations essentially related to velocity and reaction vertex uncertainties have been significantly reduced for in-beam $\gamma$-ray spectroscopy by the development of the MINOS system~\cite{obertelli14,Santamaria18} mentioned previously. MINOS was designed to fit within the DALI2 array at the RIBF. Compared to the standard passive heavy-ion target, the application of the MINOS setup led to (i) a significant increase in luminosity thanks to the use of a thick LH$_{2}$ target with possible thicknesses of 50,100 and 150 mm (corresponding to a maximal gain in statistics of a factor of $\sim$5) and (ii) a reduction of the velocity uncertainty for the Doppler correction of $\Delta\beta$/$\beta$ (from approximately 7.0\% to 0.3\%). A scientific program aiming at a systematic search for new 2$^{+}_{1}$ energies from $^{52}$Ar to $^{110}$Zr has been successfully conducted at the RIBF with MINOS (see for examples~\cite{liu2019,Chen2019,taniuchi2019,Santamaria2015,Flavigny2017,Chen2017,Lettmann2017,Shand2017,Oliver2017,Paul2017,Paul2019,Cortes2020,Sun2020,Frotscher2020,Browne2021}). 

Nevertheless, the MINOS+DALI2 setup still suffered from a rather moderate $\gamma$-ray energy resolution after Doppler correction ($\simeq10\%$ in FWHM at 1 MeV and $\beta=0.6$). 
The velocity uncertainty improvement did not lead to a significant improvement in the total energy resolution because the angular resolution coming from the size of the NaI(Tl) crystal of DALI2 and their intrinsic energy resolution still dominated. Such an energy resolution strongly limits its ability to disentangle close-lying photo peaks, typically $<$300 keV, frequently appearing in $\gamma$ spectra of odd-even exotic nuclei or exotic nuclei with heavier masses sometimes well-deformed exhibiting rotational bands or subject to shape-coexistence.

The granularity and intrinsic resolution limitations are expected to be tackled by the new-generation high-resolution $\gamma$-ray tracking arrays, such as the European AGATA spectrometer \cite{agata} and the American GRETA array \cite{greta}. AGATA and GRETA offer not only a high $\gamma$-ray efficiency and an excellent intrinsic energy resolution (0.2\% FWHM), but also an unprecedented position resolution of 4 mm. Figure\,\ref{gamma_resolution}(c) shows that the combination of such Germanium tracking detectors with a thick LH$_{2}$ target and a proton tracker for ($\emph{p}$, 2$\emph{p}$) reactions can lead to resolution down to 1.5--2\% (FWHM). Even if a significant loss in luminosity is to be expected due to the lower efficiency of germanium arrays compared to scintillator arrays, this gain in resolution could be quite useful when close-lying states of interest are populated in the nuclei of interest. However, given that the energy resolution becomes clearly dominated by the angular uncertainty, it is important that the reaction vertex resolution remains below the typical position resolution (4$\sim$5 mm in FWHM) of $\gamma$-tracking arrays to preserve this benefit.

The design of the STRASSE tracker described in the next sections aims at reaching this next step in high-resolution $\gamma$-ray spectroscopy, and consequently requires a vertex resolution better than 4\,mm (FWHM) while remaining as compact as possible to maximize $\gamma$-ray detection efficiency.\\

\subsection{Interest of combining particle and $\gamma$ spectroscopy for QFS studies}

Finally, the combination of both techniques mentioned previously in a single given measurement, namely in-beam $\gamma$-ray spectroscopy and missing mass measurements with a decent energy resolution ($\sigma_{E_x}<$\,2\, MeV), would allow to determine the absolute excitation energies of populated states and consequently enhance our capabilities to build complex level schemes of exotic nuclei for bound and unbound states. Additionally, missing-mass spectroscopy provides a unique way to measure the bound states which do not decay by prompt-$\gamma$ transitions as well as highly excited states that decay via multi-neutron emission. For example, for nuclei lying inside the so-called Island of Inversion, particle-hole "intruder" configurations (np-nh) driven by the multipole correlations are energetically favored compared to the normal shell model configuration (0p-0h), leading to the possible coexistence of a deformed ground state ($0_{1}^{+}$) and a spherical excited state ($0_{2}^{+}$). The location of the $0_{2}^{+}$ state offers a unique way to uncover the competition between shell gaps and correlations, and will help to gain new insight into the nuclear forces. However, it is sometimes experimentally challenging to measure the excited $0^{+}$ states, since they can lie very close to the 2$^{+}_{1}$ state or at lower excitation energy and could not decay via prompt $\gamma$ transitions \cite{Wimmer2009,Wimmer2010}. A missing-mass measurement from particle detection is thus often required to pin down their excitation energies. In case of the unbound states in extreme neutron-rich nuclei such as $^{28}$O and $^{10}$He, the reaction $^{29}$F$(p,~2p)^{28}$O and $^{11}$Li$(p,~2p)^{10}$He in inverse kinematics could be performed. By measuring the two recoil protons, together with the incident particle, the excitation energy of the residue as well as the momentum of the struck proton, namely, the missing mass and missing momentum can be reconstructed. In this case, we do not need any information regarding the decay of the residue including those four neutrons in $^{28}$O and six neutrons for the 6n decay states in $^{10}$He. The coincidence measurement of multi-neutrons suffers from extremely low efficiency and makes invariant mass spectroscopy either difficult or impossible.\\

%% file: Simulations.tex
\begin{figure}[htp]
\begin{center}
\includegraphics[width=7cm]{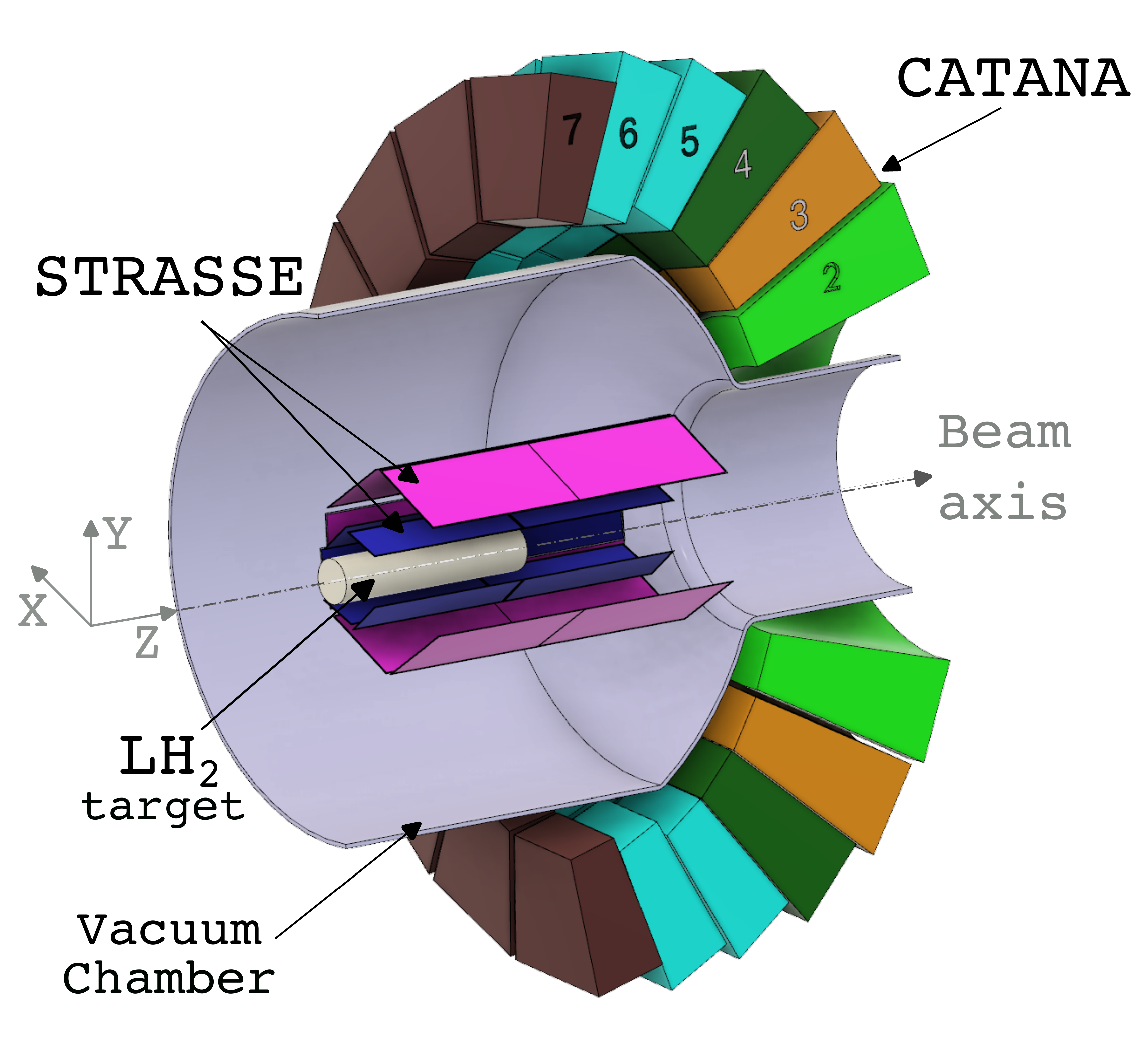}
\caption{\label{strasse} Schematic representation of STRASSE (silicon tracker: inner barrel (blue), outer barrel (pink)) with its 150-mm thick LH$_{2}$ target (gray)  and the CATANA array (layers 2 to 7).}
\label{strasse_scheme}
\end{center}
\end{figure}

Given the importance of angular and vertex resolution to reach the targeted sensitivity ($\sigma_{E_x}<$\,2\,MeV, $\sigma_{vertex}<$\,1\,mm, see Section\,\ref{sec_intro}), it was decided that the STRASSE proton tracker will be made of an array of highly-segmented silicon detectors (double-sided silicon strip detectors, DSSD) in vacuum. This technology is used for most charged particle trackers requiring a very high granularity while maintaining a small material budget to minimize straggling. Since STRASSE is intended to be used with the CATANA array of CsI(Na) scintillators\,\cite{catana} for missing mass studies, a schematic view of their coupling with the LH$_{2}$ target cell is displayed in Fig.\,\ref{strasse_scheme} for illustration.

In this section, we will give details about the design choices made for the proton tracker (geometry, segmentation, etc) and for the LH$_{2}$ target based on Monte-Carlo simulations performed and analyzed coherently with the $\emph{nptool}$\,\cite{nptool} package (relying on Geant4~\cite{geant4} and ROOT~\cite{root} toolkits).

\subsection{Kinematics of quasi-free scattering reactions $(p, 2p)$ and $(p, 3p)$}

\begin{figure*}[htpb!]
\begin{center}
\includegraphics[width=0.8\textwidth]{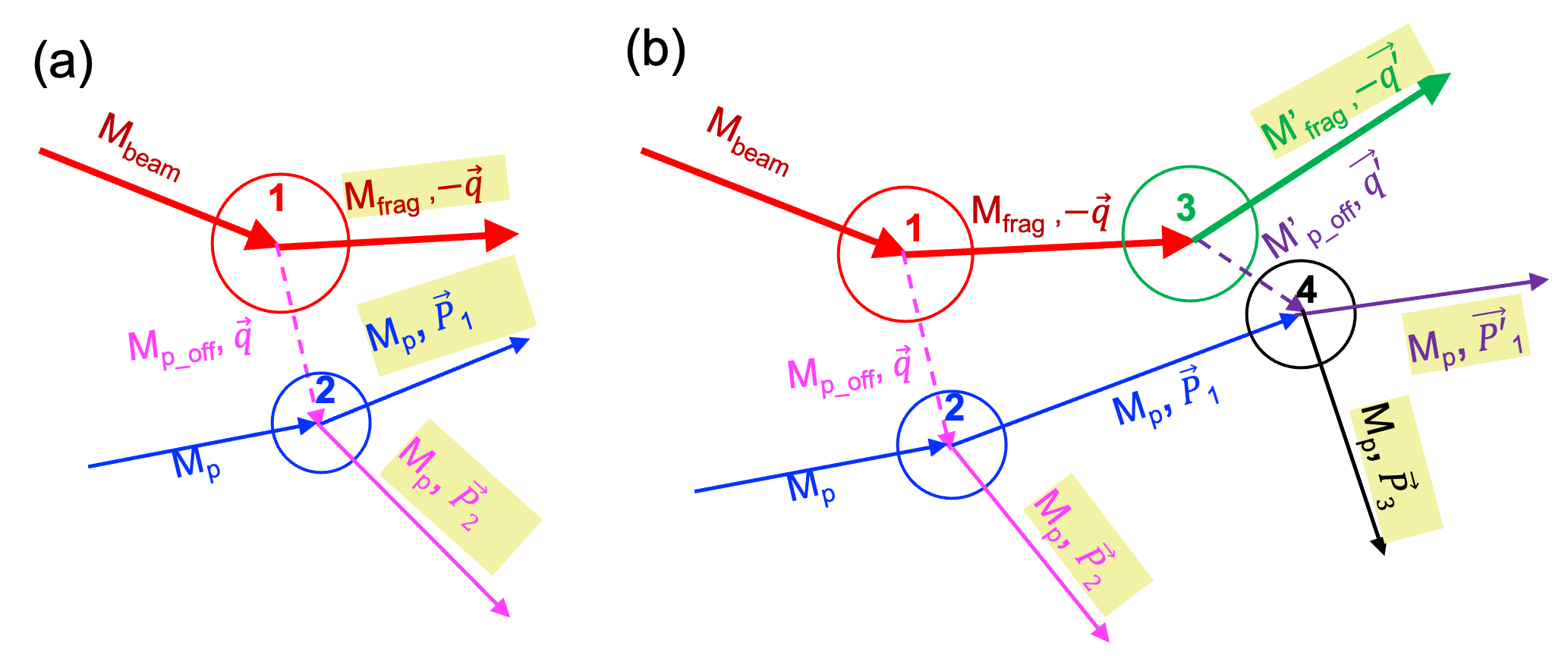}
\caption{\label{p2p_diagram} The Feynman diagrams to describe the reaction process of $(p, 2p)$ (a) and $(p, 3p)$ (b).}
\end{center}
\end{figure*}

\subsubsection{$(p, 2p)$ reactions}
An event generator for QFS reactions was implemented in the \emph{nptool} framework following a similar approach to Ref.~\cite{Chulkov2005,Panin2016} within the impulse approximation. The Feynman diagram used to compute the kinematics of the particles involved in quasi-free $(p, 2p)$ reactions is shown in Fig.\,\ref{p2p_diagram}(a). The proton target interacts with the projectile with a mass of $M_{\text{beam}}$, knocking out a proton with an internal momentum $\vec{q}$ and leaving behind a fragment with the mass of $M_{\text{frag}}$. The reaction is described in two steps, represented by two vertices at which energy and momentum conservation is fulfilled:
\begin{itemize}
    \item Vertex 1 corresponds to the dissociation of the projectile into the fragment and an intermediate virtual proton with an off-shell mass M$_{p_\text{-off}}$. This off-shell mass is deduced from energy conservation in the rest frame of the projectile.
 
    \item Vertex 2 denotes the elastic scattering process between the intermediate virtual proton and the target proton, leaving two outgoing protons with their physical mass $M_{p}$ after the reaction. 
    Lorentz transformation is used to convert the momentum of the virtual proton in the projectile rest frame ($\vec{q}$) to the laboratory frame ($\vec{P}$). 
   
    Then the invariant mass (S) of the two-body system including the virtual proton and the proton target is calculated using
    \begin{equation}
    S^2 = M_{p_\text{-off}}^2+M_{p}^2+2M_{p}\sqrt{\vec{P}^2+M_{p_\text{-off}}^2},
    \label{s_inv}
    \end{equation} 
where M$_{p}$ is the physical mass of the proton. 
\end{itemize}
With the information of S, $M_{p_\text{-off}}$ and $M_{p}$, the kinematics of the two recoil proton is then calculated in the center of mass frame assuming an isotropic scattering process. It should be noted that the distribution of intrinsic momentum for the removed nucleon in the projectile ($\vec{q}$) is a free parameter and can be estimated using the Goldhaber model\,\cite{Gold1974} or extracted from experimental data. 

\begin{figure*}[htpb!]
\begin{center}
\includegraphics[width=0.95\textwidth]{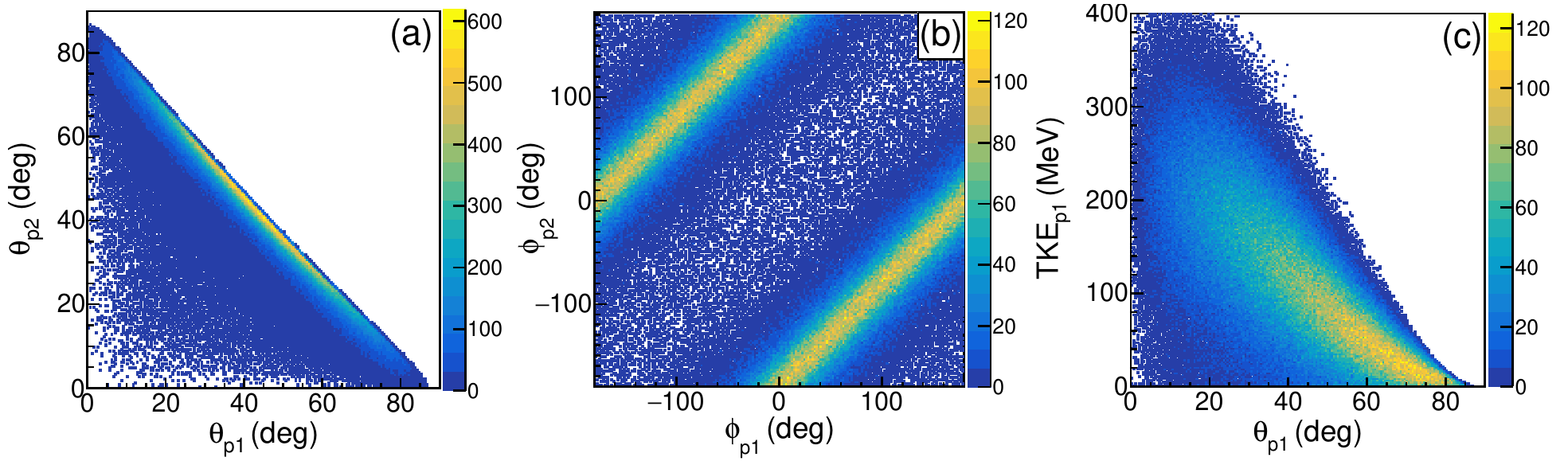}
\caption{\label{p2pkine} Simulated angular distributions of the two recoil protons from the $^{17}$F$(p, 2p)^{16}$O reaction at 250 MeV/nucleon. Correlations between polar ($\theta$) and azimuthal ($\phi$) angles are shown in panels (a) and (b). Correlations between the polar angle and the kinetic energy of the proton are shown in panel (c).}
\vspace{0.2cm}
\includegraphics[width=0.98\textwidth]{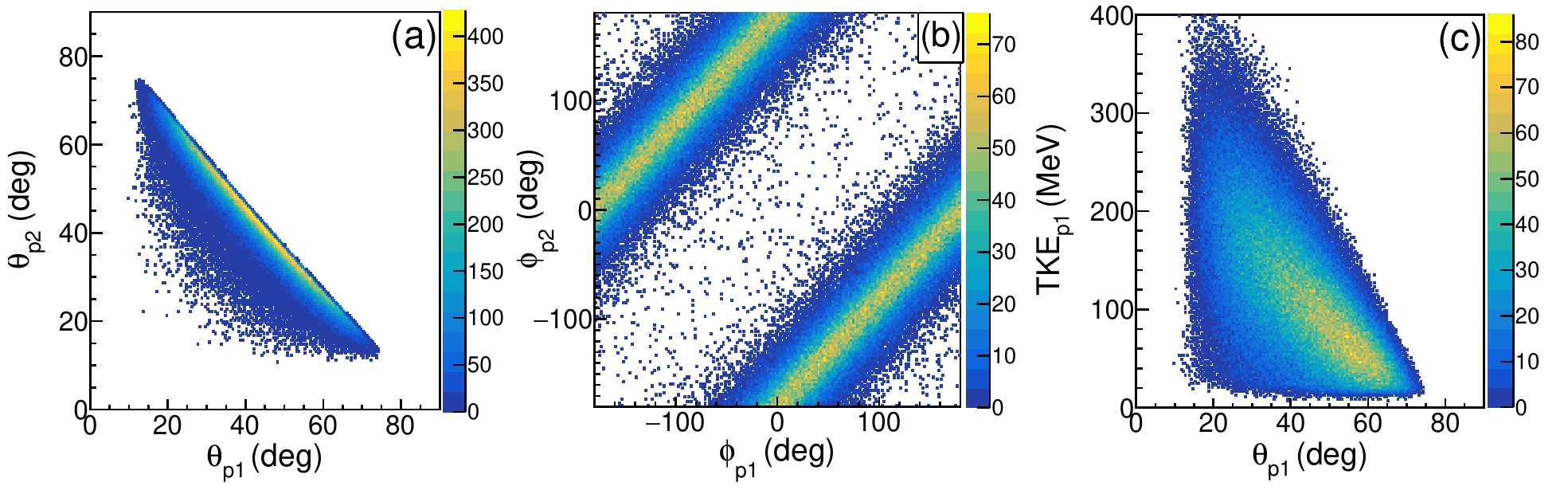}
\caption{\label{p2pkine_det}Kinematic variables of recoil protons (same as Fig.\,\ref{p2pkine}) requiring that the two outgoing protons interact in the active area of both silicon layers.}
\end{center}
\end{figure*}

For all the simulations presented throughout this document, if not specified otherwise,  we used a set of events generated for the $^{17}$F$(p, 2p)^{16}$O reaction at 250 MeV/nucleon with a typical beam size of $\sigma_x =$\,6\,mm, $\sigma_y =$\,4\,mm. The momentum of the struck nucleon in the projectile rest frame was considered to be randomly distributed within a gaussian of 100 MeV/c width (FWHM) in all directions. The angular distributions of the two recoil protons simulated are displayed in Fig.\,\ref{p2pkine}(a,b) and their kinetic energy as a function of angle in Fig.\,\ref{p2pkine}(c). These correlations are similar to what is observed in proton elastic scattering reactions, but smoothed by the internal motion of the knocked-out proton inside the nucleus and affected by the binding energy of the knocked-out proton.\\ 

\begin{figure*}[htpb!]
\begin{center}
\includegraphics[width=0.75\textwidth,trim={0.2cm 0.2cm 0.2cm 0.2cm},clip]
{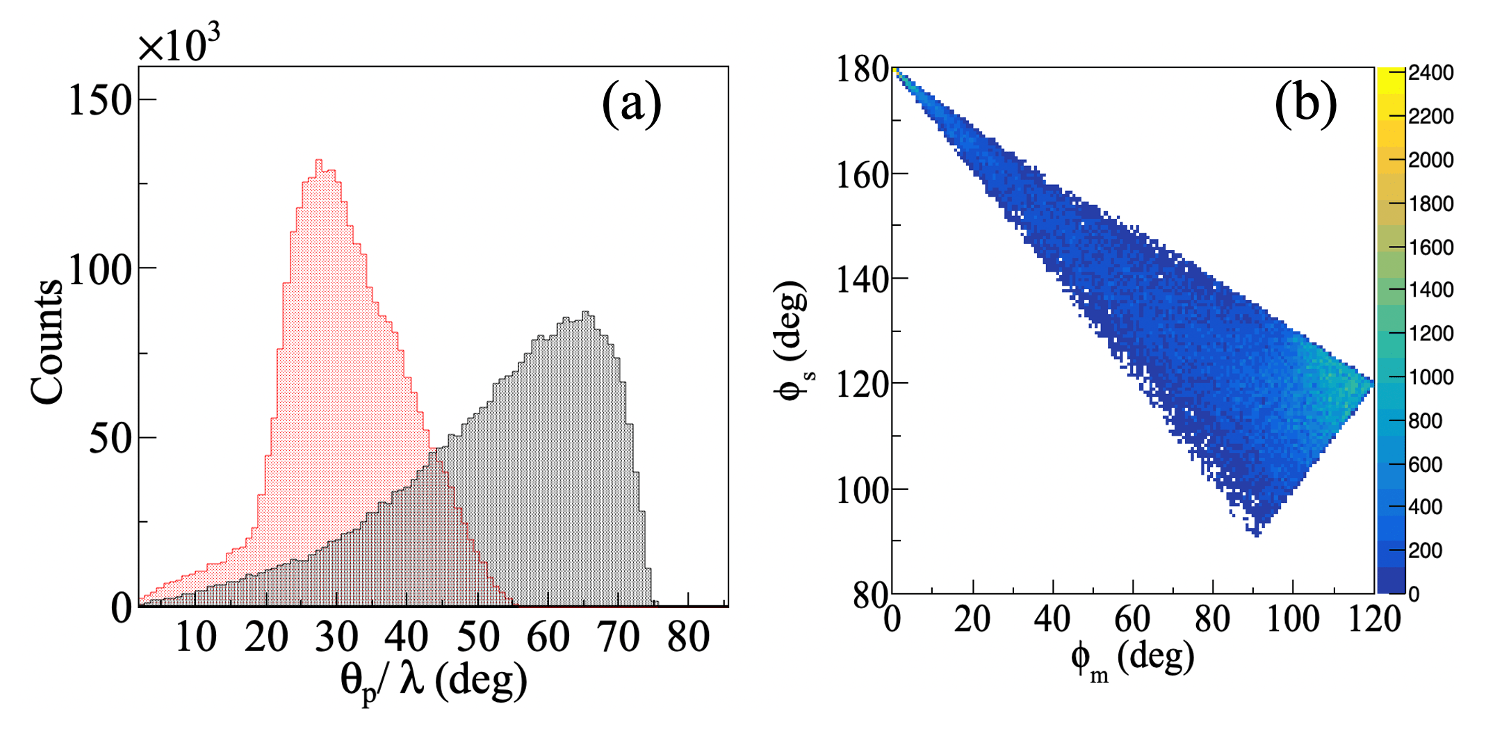}
\caption{\label{p3p_kine} Simulated angular distributions of three recoil protons from the $^{56}$Ti$(p, 3p)^{54}$Ca reaction at 210\,MeV/nucleon: (a) Distribution of the polar angle $\theta_{p}$ between each outgoing proton and the beam axis (red) and the opening angle $\lambda$ between each pair of scattered protons (black); (b) Correlation between the two smaller azimuthal angles $\phi_{s}$ and $\phi_{m}$ projected in the plane perpendicular to the beam axis. }
\end{center}
\end{figure*}

\subsubsection{($p, 3p)$ reactions}

In several quasi-free scattering studies, the $(p, 3p)$ reaction mechanism was also found rather useful for spectroscopic studies~\cite{taniuchi2019} partly due to its different final state selectivity compared to $(p, 2p)$. In this perspective, the STRASSE system also targets the study of this promising reaction channel involving the detection of three outgoing protons. 

From data taken at the RIBF at $\sim$ 250 MeV/nucleon in the mass region A=68$\sim$122~\cite{Frotscher2020}, this $(p, 3p)$ reaction mechanism has been shown to be a two-step direct process in which the two protons are sequentially removed from the projectile, i.e. two sequential $(p, 2p)$ processes. As shown in Fig.\,\ref{p2p_diagram}\,(b), the $(p, 3p)$ reaction are presented by four vertices. Vertex 1 and 2 correspond to the first proton removal process from the projectile, exactly the same as the $(p, 2p)$ process shown in Fig.\,\ref{p2p_diagram}(a). At Vertex 3, $M_{frag}$ from Vertex 1 dissociates into the final fragment of mass $M^{'}_{frag}$ and another intermediate virtual proton with the off-shell mass of $M^{'}_{p_-off}$. At Vertex 4, one of the recoil protons from Vertex 2 further interacts with $M^{'}_{p_-off}$. The resulting three recoil protons and the outgoing fragment from $(p, 3p)$ are denoted with the yellow shaded area in Fig.\,\ref{p2p_diagram} (b). Same as the denotation used in ~\cite{Frotscher2020}, $\theta$ is the polar angle of recoil protons, $\lambda$ is the angle between each pair of protons, and $\phi$ denotes the angles between the protons in the plane perpendicular to the beam axis ($\phi_{s} \leq \phi_{m} \leq \phi_{l}$).  Fig.\,\ref{p3p_kine} shows the simulated angular distributions of the three recoil protons from the $^{56}$Ti$(p, 3p)^{54}$Ca reaction at 210\,MeV/nucleon. 
Different from the $(p, 2p)$ reactions in which the $\theta$ distribution peaks at $\sim$ 40 degrees and the $\lambda$ distribution peaks at $\sim$ 80 degrees, the peak value of $\theta$ and $\lambda$ in $(p, 3p)$ both shift towards a lower angle as shown in Fig.\,\ref{p3p_kine} (a). The correlation between $\phi_{s}$ and $\phi_{m}$ shown in Fig.\,\ref{p3p_kine} (b) is similar to what observed in $^{81}$Ga$(p, 3p)$ reaction in ~\cite{Frotscher2020}.

\subsection{Target geometry}

The optimum target thickness, that maximizes the ($\emph{p}$, 2$\emph{p}$) reaction rate without multiple reactions in the target at the RIBF beam energy for medium mass nuclei, has been calculated to be in the 100--150 mm range \cite{obertelli14}. While for heavy isotopes like Z = 50, the optimum target thickness will be $\sim$ 50 mm due to the large energy loss. 150\,mm is adopted as a standard maximum for the STRASSE system. As mentioned previously, the radius of the target on one hand has to be small to minimize the angular straggling of the recoil protons impacting directly the missing-mass resolution, while on the other hand it has to be large to cover the spread of incoming secondary beams from the in-flight separator. The typical beam profile at the secondary target position of the SAMURAI spectrometer\,\cite{SAMURAI1,SAMURAI2} is about 6\,mm ($\sigma_{x,y}$).

The technical design of the target cell for STRASSE, detailed later in Section\,\ref{sec:lh2target}, requires that the entrance window radius has to be slightly smaller than the radius of the long cylindrical part of the cell so that LH$_{2}$ intake and exhaust tubes can fit in the external ring. 

This means that a compromise between beam acceptance and straggling of outgoing protons had to be made. It was chosen to use an effective target radius of 10 mm (entrance window radius) giving a beam acceptance of $\sim$60\%. This entrance radius implies a cylindrical target cell radius of 15.5 mm leading to an angular straggling of 3.0 mrad as previously listed in Table\,\ref{tab:straggling}.  

\subsection{Tracker geometry : Efficiency}
 \label{sec:efficiency}

\begin{figure*}[htpb!]
\begin{center}
\includegraphics[width=0.7\textwidth]{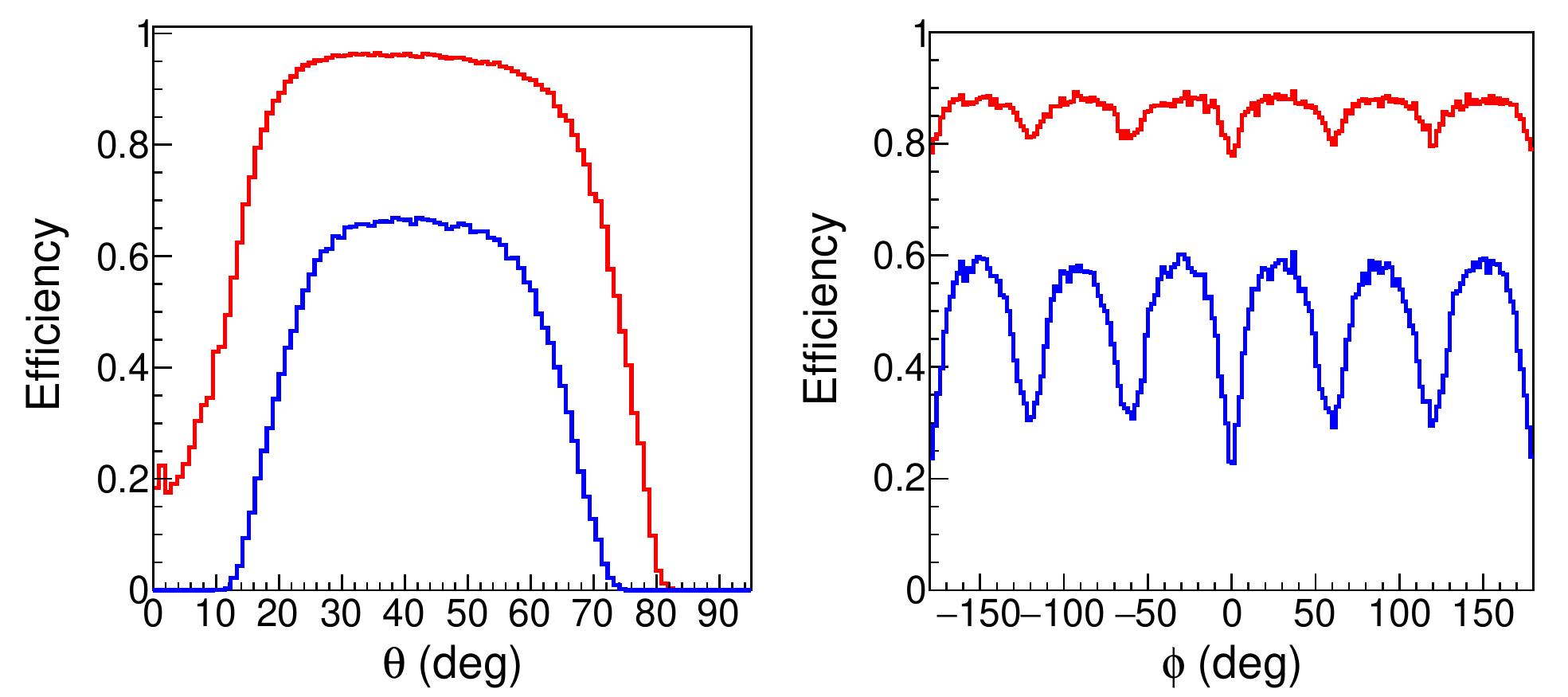}
\caption{Geometrical efficiency of crossing the active area of both silicon layers as a function of polar (Left) and azimuthal (Right) angle of emitted protons. Red curve corresponds to at least one proton interacting in two silicon layers. The blue curve corresponds to two-detected protons, each interacting in both layers.}
\label{efficicency_interaction}
\end{center}
\end{figure*}

Several constraints emerge from the need to detect most of the $(p, 2p)$ events occurring all along the cylindrical 150-mm LH$_{2}$ target with outgoing proton polar angles ($\theta_{p1}$,$\theta_{p2}$) centered around 40 degrees and isotropically distributed azimuthal angle ($\phi_{p1}$,$\phi_{p2}$) (see Fig.\,\ref{p2pkine}): 
\begin{itemize}
    \item tracking layers should be around 300-mm long (along beam axis),
    \item tracking layers must encircle the target cell to cover nearly 2$\pi$ in $\phi$ with an optimum detection efficiency,
    \item tracking layers should be offsetted along the beam axis with respect to the entrance of the target cell and between the different layers themselves (forward focused protons around 40 degrees).
\end{itemize}
 Additionally, strong technical limitations originate on the manufacturing of silicon sensors and the coupling with CATANA and the LH$_{2}$ target must be considered:
\begin{itemize}
    \item individual silicon detectors are limited in size due to the maximum wafer size (6 inches, 15.24\,cm),
    \item a minimal radius of approximately 20 mm is imposed by the LH$_{2}$ cell external radius,
    \item the mechanical structure for the tracking layers should minimize the efficiency losses,
    \item the full silicon array and its front-end electronics must fit in a cylinder of 180 mm radius (defined by CATANA).
\end{itemize}

Based on these constraints, the STRASSE silicon array was designed in a hexagonal configuration with two tracking layers (inner barrel and outer barrel) displayed in Fig.\,\ref{strasse_scheme}. Each layer/barrel is composed of 6 assembly of 2 DSSDs chained along the beam axis to cover a length of approximately 240 mm. The inner and outer detectors are placed at 30 and 61 mm from the beam axis, respectively.

The resulting kinematical coverage of this geometry (no electronic threshold at this stage of the discussion) is displayed in Fig.\,\ref{p2pkine_det}. Requiring that both protons cross the active silicon area leads to a polar angular coverage $\theta_{p}$ between approximately 15 and 75 degrees and a mostly flat coverage of the azimuthal angles. Protons below $\sim$15\,MeV are undetected simply because they stop in the radial thickness of the target (see Fig.\,\ref{p2pkine_det}(c)). The resulting efficiencies are displayed as a function of angles in Fig.\,\ref{efficicency_interaction} using the emitted protons from Fig.\,\ref{p2pkine} as a reference. The integrated one- and two-proton efficiency are 86\% and 49\% respectively for this set of events. It is worth noting that the momentum spread of the struck nucleon ($\sigma_{0}$) considered in the input broadens the proton kinematic in energy and angle and thus reduces to some extent the two-proton integrated efficiency (see Table\,\ref{tab:efficiencygeo}).

\begin{table}[htp]
  \renewcommand\thetable{2}
    \centering
        \caption{Proton efficiency to cross the active area of both silicon layers for different simulation inputs: varying momentum widths of the knocked-out proton, using a perfect pencil beam (no position/angular spread on target) or using a realistic beam as specified in the text.}
    \label{tab:efficiencygeo}
    \begin{tabular}{c|ccc|ccc}
    \hline \hline
        & \multicolumn{3}{c}{Pencil beam} & \multicolumn{3}{|c}{Realistic beam} \\
    \hline
     $\sigma_{0}$(MeV/c) & 0 & 50 & 100 & 0 & 50 & \textbf{100} \\
     \hline
     $\geq$1p efficiency (\%) & 79 & 85 & 87 & 84 & 85 & \textbf{86} \\
     2p efficiency (\%)       & 65 & 59 & 55 & 53 & 52 & \textbf{49} \\
    \hline \hline
    \end{tabular}
\end{table}

\subsection{Tracker thickness and granularity : Missing mass and vertex resolution }

\begin{figure}[htpb!]
  \begin{center}
     \includegraphics[width=0.47\textwidth]{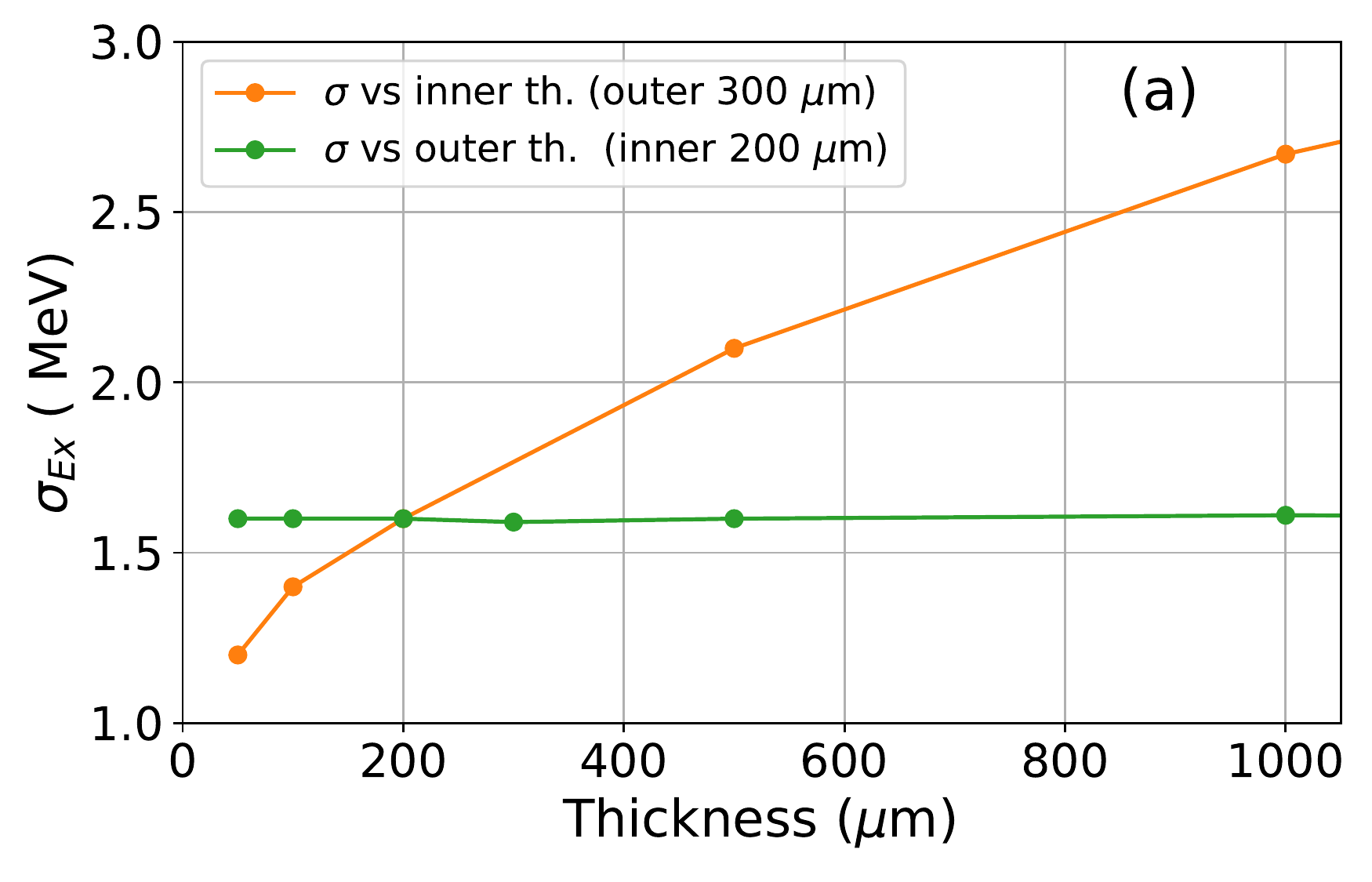}
     \includegraphics[width=0.47\textwidth]{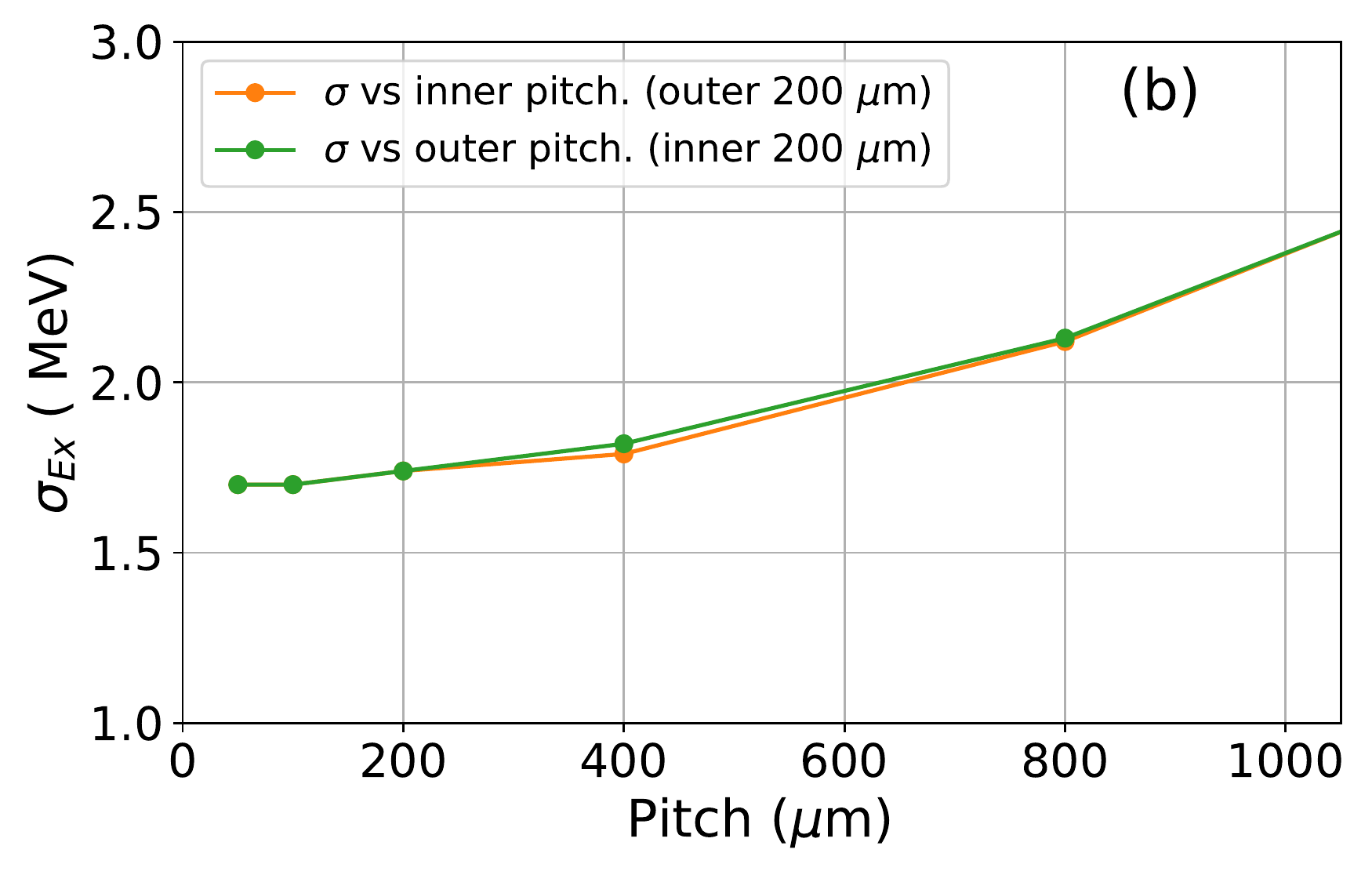}
    \caption{(a) Missing mass resolution as a function of the thickness of each silicon layer (varying inner in orange, outer in green). The strip pitch of inner and outer layers are fixed to 200\,$\upmu$m. (b) Missing mass resolution as a function of the strip pitch of each silicon layer for constant thicknesses (inner/outer: 200/300\,$\upmu$m). In all cases the LH$_{2}$ target is 150-mm thick with a radius of 15.5 mm but the constant Mylar window contribution to the resolution is not included.}
      \label{fig:MMres}
  \end{center}
\end{figure} 

\begin{figure}[htpb!]
  \begin{center}
      \includegraphics[width=0.47
\textwidth]{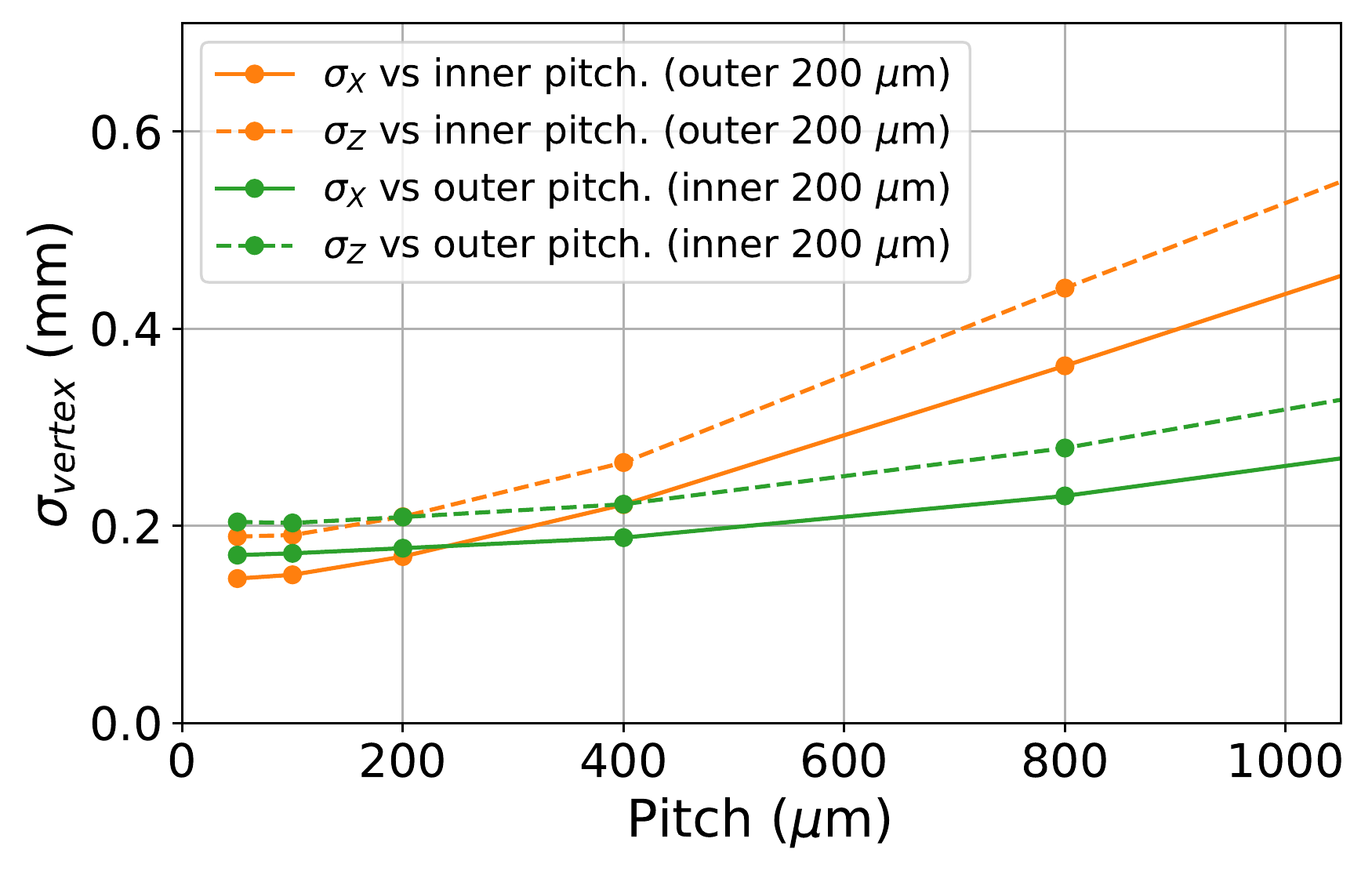}
    \caption{Vertex resolution as a function of the pitch of each silicon layer (varying inner in orange, outer in green). The solid line corresponds to the vertex resolution in the plane perpendicular to the beam axis (X,Y coordinates) and the dashed line along the beam axis (Z coordinate). The thicknesses of inner and outer layers are kept constant to 200\,$\upmu$m and 300\,$\upmu$m, respectively.}
      \label{fig:vertexres}
  \end{center}
\end{figure} 

As mentioned in Section~\ref{sec:mmreq}, the dominating factor for the missing mass resolution is the angular resolution of the system. This angular resolution results essentially from the angular straggling in the detector material (silicon thickness)
and the granularity of the detector (strip pitch).

For the nominal target design (150-mm thick, radius of 15.5 mm), we studied the dependence of the missing mass resolution as a function of these two parameters (silicon thickness and pitch) by performing a large set of simulations in which they vary. These simulations were analysed consistently following three main steps: (1) threshold and energy matching, (2) track analysis and (3) missing-mass reconstruction. For step (1) we set a detection threshold and an energy matching condition between both sides of each silicon to allow most of the physical events to be detected according to the thickness chosen. For step (2) we keep only events with two protons detected, calculate their tracks based on the interaction points in the inner/outer layers and determine the reaction vertex as the middle of the segment linking the two tracks which minimizes the distance between them, same as method used for MINOS \cite{Santamaria18}. Step (3) is simply a missing-mass calculation assuming at this stage that the total kinetic energy of protons can be measured with a resolution of 1.7\% (FWHM) from recent CATANA tests but without a true simulation of the CATANA calorimeter detection. The results are displayed in Fig.\,\ref{fig:MMres} and Fig.\,\ref{fig:vertexres}. 

As anticipated, we can see in Fig.\,\ref{fig:MMres}(a) that the missing mass resolution is almost proportional to the thickness of the inner layer due to the angular straggling in the detector but barely depends on the outer layer thickness. To reach the targeted resolution $\le$ 2\,MeV it is necessary to stay below an inner layer thickness of 400\,$\upmu$m. While a thinner inner layer is definitely preferable for resolution purposes, it is more constraining in terms of fragility but also electronic noise requirement because the energy deposit of protons diminishes significantly. More details about energy deposit considerations will be given in the next section.

Concerning the strip pitch, the missing mass resolution displayed in Fig.\,\ref{fig:MMres}(b) shows that below pitches of 200-300\,$\upmu$m, a plateau around $\sigma_{E_x} =1.7$\,MeV is reached and there is little gain in reducing the strip size. It can be understood because the angular straggling due to the inner detector thickness becomes the main contribution to the total angular resolution. Similarly, the vertex resolution shown in Fig.\,\ref{fig:vertexres} converges around $\sigma_{vertex}=0.17$\,mm for pitches below 200-300\,$\upmu$m and is more sensitive to the pitch of the inner layer than the outer layer.\\

\subsection{Energy deposit and threshold}

\label{sec:penergyloss}
\begin{figure*}[t!]
\begin{center}
\includegraphics[width=0.47\textwidth]{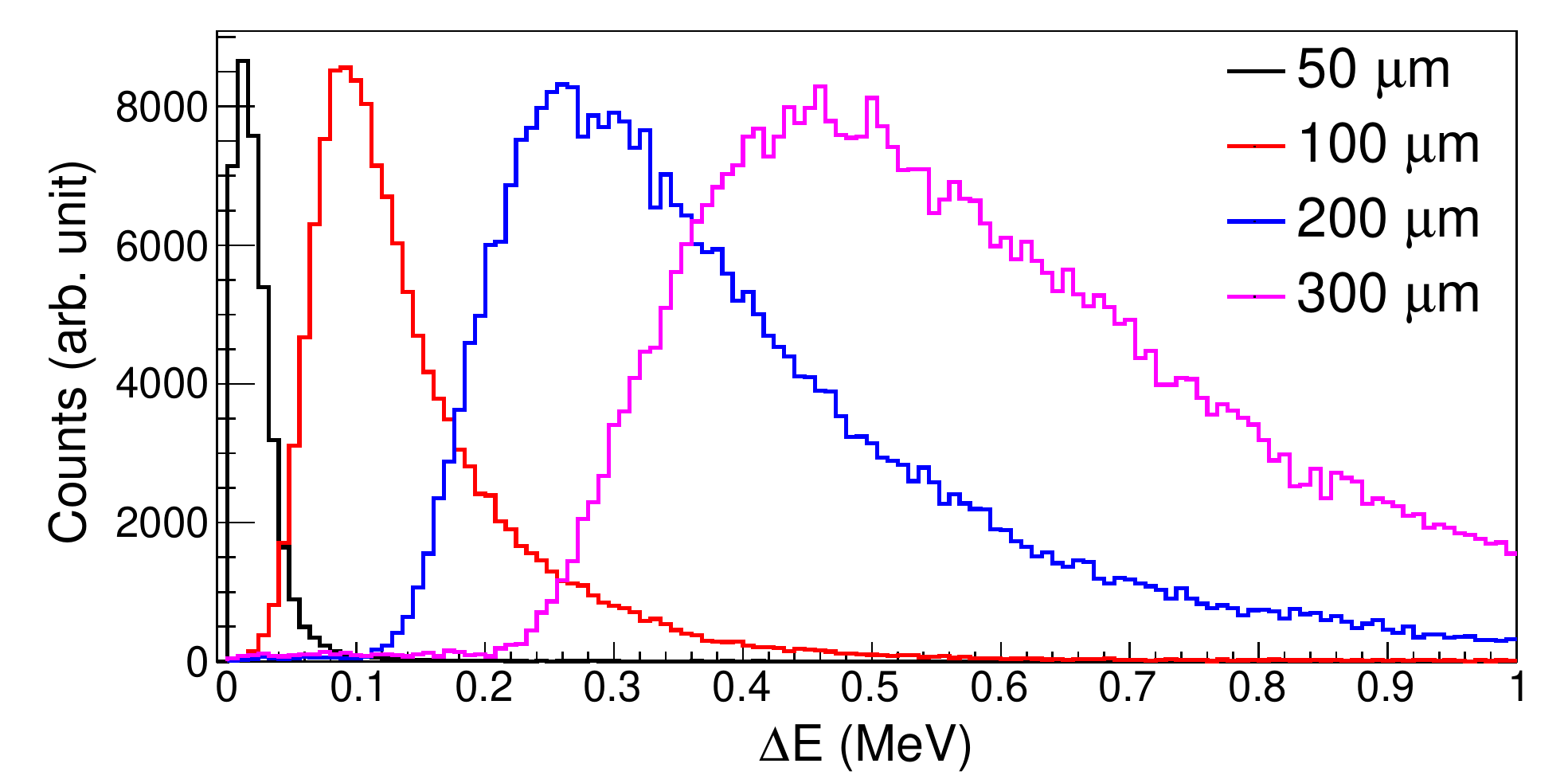}
\includegraphics[width=0.43\textwidth,trim={0 0.8cm 0 0.5cm},clip]{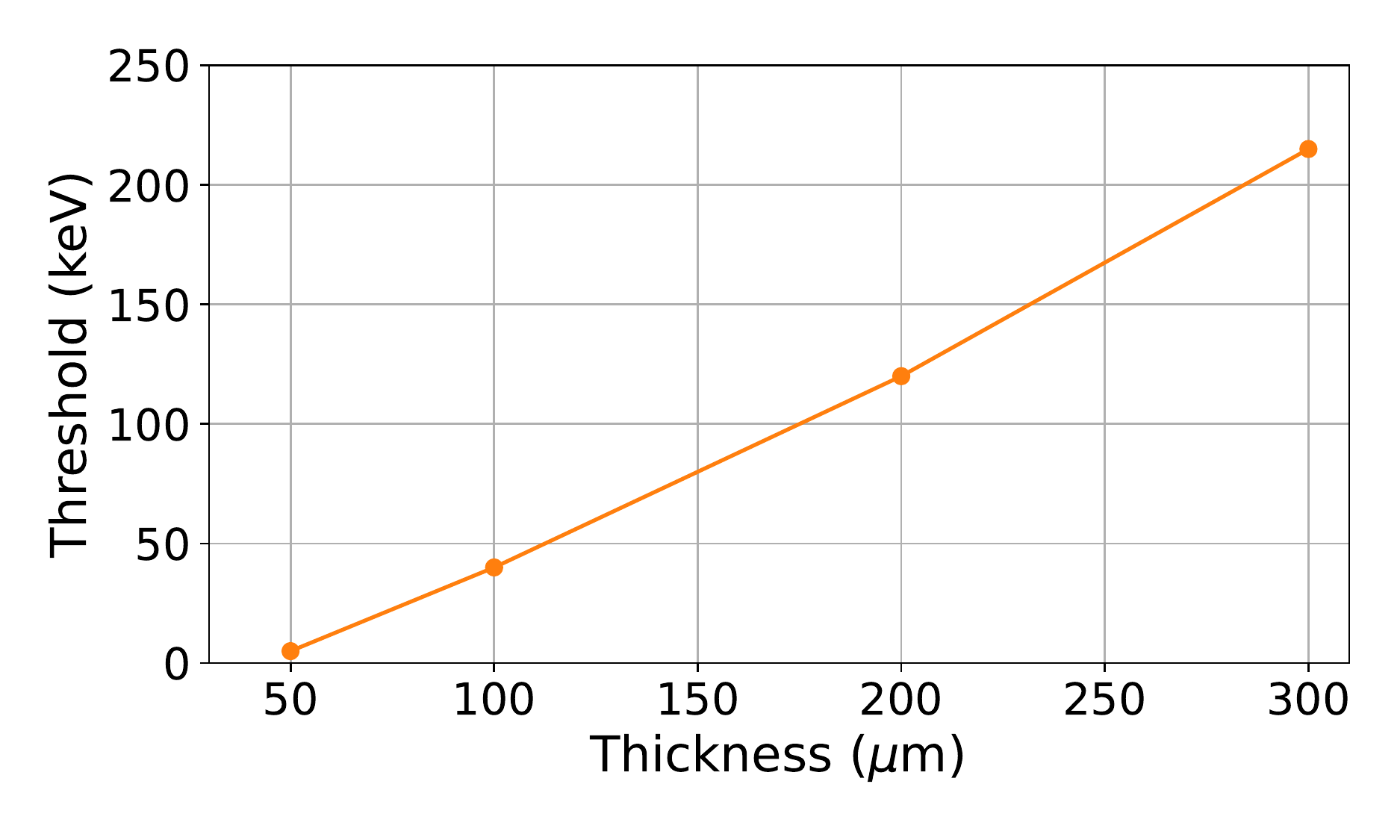}
\caption{ (Left) Energy deposited by protons in the inner detection layer for different silicon thicknesses. The distributions have been arbitrarily normalized for shape comparison. (Right) Required energy threshold to detect most of the protons (95\%) for a given inner layer thickness.}
\label{fig:penergyloss}
\end{center}
\end{figure*}

\begin{figure*}[htpb]
\begin{center}
\includegraphics[width=0.8\textwidth]{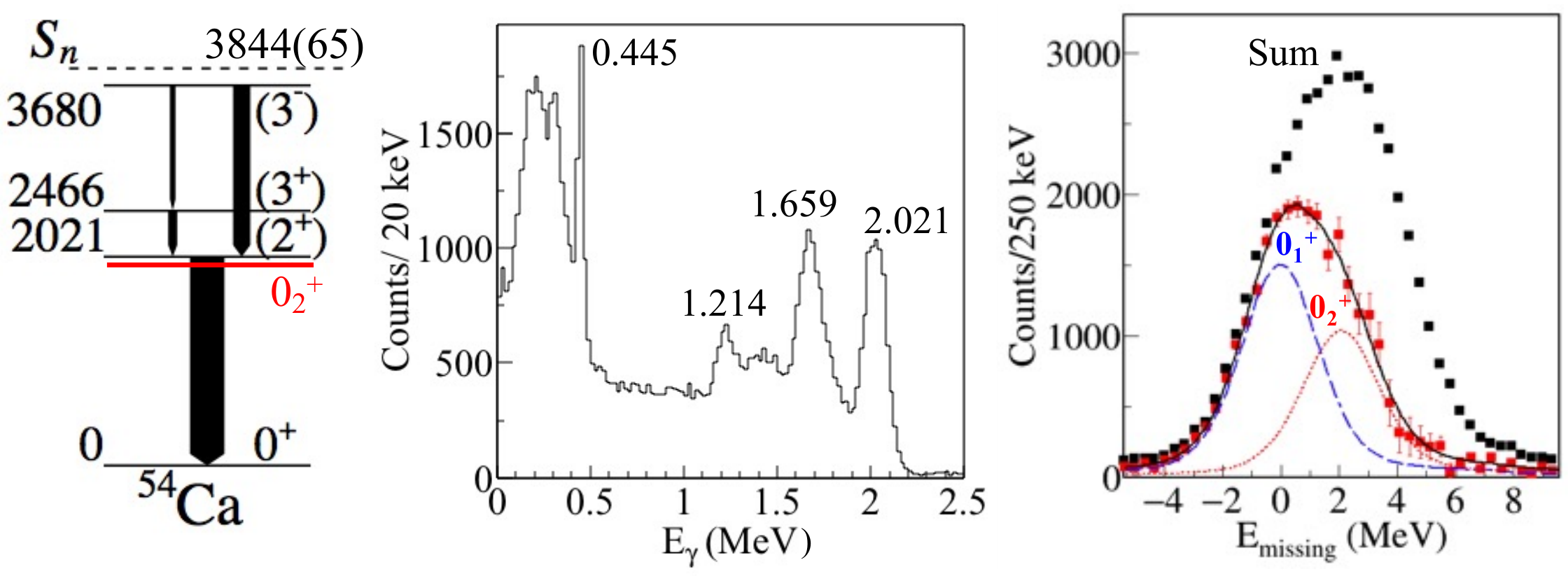}
\caption{\label{scenario1} (Left) Level scheme of $^{54}$Ca in a scenario with a 0$_{2}^{+}$ state slightly below or very close to the 2$_{1}^{+}$ state. (Center) Simulated gamma spectrum of CATANA. (Right) Simulated missing-mass spectra from $^{56}$Ti$(p, 3p)^{54}$Ca: (black squares) total spectrum, (red squares) strength from the ground 0$^{+}_{1}$ and isomeric 0$^{+}_{2}$ state deduced by subtracting the gamma-gated spectrum (not displayed) from the total. The black solid line represents the best fit to the subtracted spectrum. The blue-dashed and red-dotted lines represent the simulated response of the setup to the ground state and the 0$^{+}_{2}$ state, respectively. Note that the statistics are computed based on the estimated beam intensity, required beam time and the cross sections predicted by the GXPF1Br interaction \cite{step13}.}
\end{center}
\end{figure*}

Although diminishing the inner layer thickness was shown to be preferable for resolution purposes, there is a trade-off to make to get sufficient energy deposit in the detector compared to its noise level. As an illustration, the energy deposited by outgoing protons from the $^{17}$F$(p, 2p)^{16}$O reaction at 250 MeV/nucleon is shown in Fig.\,\ref{fig:penergyloss}\,(left) for different thicknesses of the inner barrel layer. From 300\,to 50\,$\upmu$m, the centroid of the energy loss distribution evolves from 450\,keV to only 20\,keV. From these distributions, the threshold above which 95\% of all protons can be detected is shown in Fig.\,\ref{fig:penergyloss}\,(right).  Assuming roughly that the total noise level $\sigma_{\text{ENC}}$ must be about 10 times smaller than these values to safely distinguish between noise and protons, it becomes clear that thicknesses below 100\,$\upmu$m would impose noise levels ($\sigma_{\text{ENC}}\leq$ 4 keV) which is very difficult to be reached technically given the strip capacitance. This consideration added to the fragility of such thin detectors, favored a thickness of 200 $\upmu$m for the inner layer.

\subsection{Intermediate conclusion: Tracker configuration chosen}

As a compromise between all the aspects of the simulation study discussed in the previous sections, we chose for the silicon tracker:
\begin{itemize}
\item Thicknesses of 200 and 300\,$\upmu$m for the inner and outer layers, respectively.
\item A strip pitch of 200\,$\upmu$m for both silicon layers. 
\end{itemize}
This configuration respects the targeted missing mass resolution $\sigma_{E_x}\leq 2$\,MeV and should lead to a vertex resolution $\sigma_{vertex}\simeq 0.2$\,mm (FWHM\,$\simeq 0.5$\,mm). Such a small vertex resolution has negligible contribution to the energy resolution of the $\gamma$-ray spectrum measured by $\gamma$-ray tracking array (AGATA/GRETINA) which have a $\sim$\,4-mm position resolution in FWHM for the first interaction position \cite{agata}. The chosen pitch of 200\,$\upmu$m results in 17358 readout channels for the whole silicon tracker. This number of channels is a challenge given the limited space available but allows a future margin of improvement in missing-mass resolution if an upgrade is made with a thinner inner barrel of 100\,$\upmu$m.\\

\subsection{Realistic example of missing-mass spectroscopy}

\begin{figure*}[htpb!]
\centering
\includegraphics[width=0.97\textwidth,trim={0 0 0.1cm 0},clip]{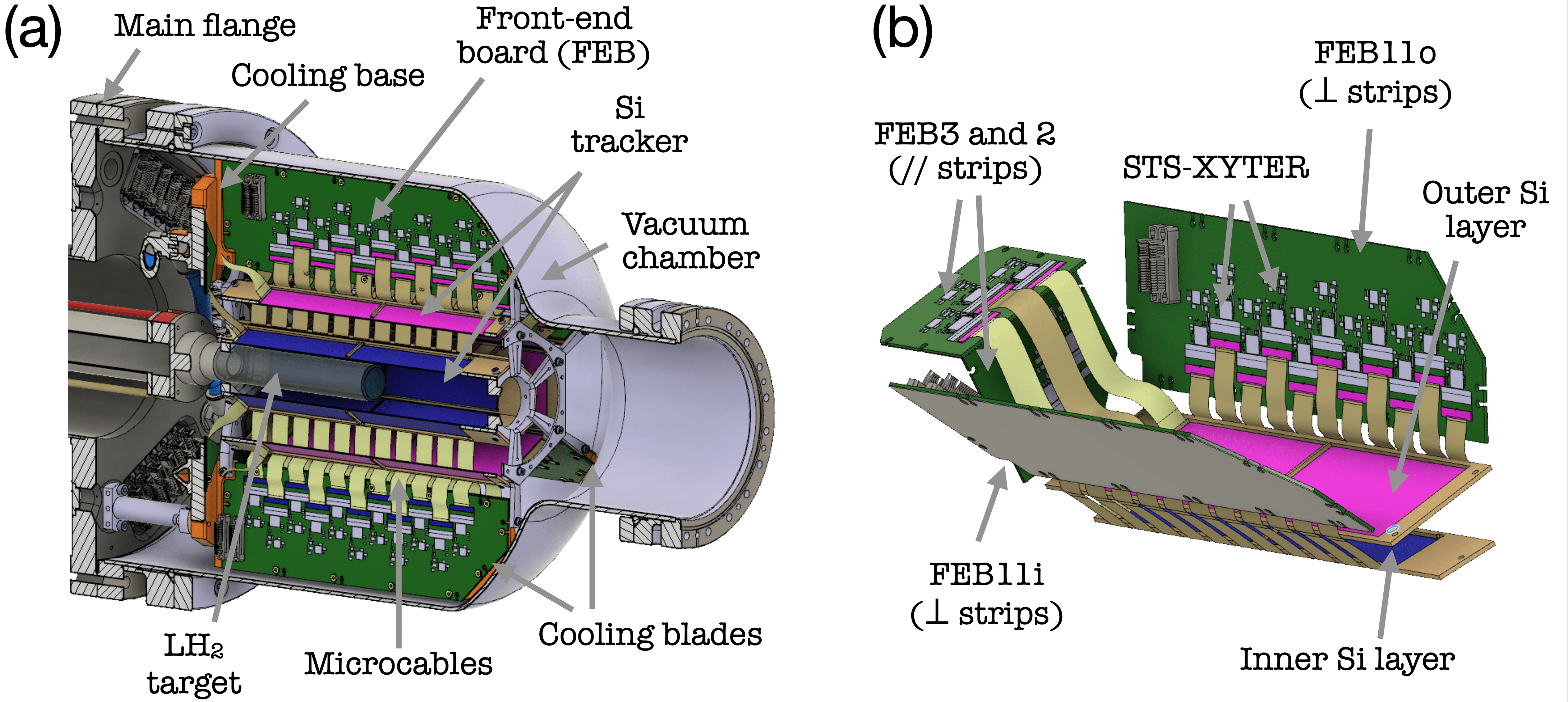}
\caption{(a) Cut view of the detailed STRASSE design and (b) single STRASSE detector module with its front-end readout electronics.}
\label{fig:strasse_overview}
\end{figure*}

The excitation energy of the 0$_{2}^{+}$ state in $^{54}$Ca can provide information on the gds-orbitals correlations and by consequence could constrain the predictions for the structure of $^{60}$Ca and for the dripline location of the Ca isotopes. Since the 0$_{2}^{+}$ state might not decay via prompt $\gamma$ transitions, we proposed to perform the missing-mass measurement together with the $\gamma$ spectroscopy from the $^{56}$Ti$(p, 3p)^{54}$Ca reaction at 205 MeV/u to search for this particular state at SAMURAI with STRASSE. The proposed experiment has been accepted by the NP-PAC committee. Fig.\,\ref{scenario1} (left) shows the measured level scheme of $^{54}$Ca \cite{Browne2021} in which the unknown 0$_{2}^{+}$ state is assumed to be isomeric and lies slightly below the 2$_{1}^{+}$ state. Population strengths to states decaying via prompt $\gamma$-rays can be extracted based on the $\gamma$-ray spectrum measured by CATANA as shown with simulations in Fig.\,\ref{scenario1}\,(center). The subtraction of these contributions from the inclusive missing-mass spectrum can give access to the spectroscopic strength of states not decaying by prompt $\gamma$-ray transitions, i.e., the ground 0$_{1}^{+}$ and the isomeric 0$_{2}^{+}$ state. The red squares in Fig.\,\ref{scenario1} (right) show the simulated missing-mass spectrum after this subtraction procedure and the black solid line is the best fit in which the amplitudes of two 0$^{+}$ states together with the peak position of the 0$_{2}^{+}$ state are free parameters. The blue dashed and red-dotted lines represent the simulated response of the setup to the ground state and the 0$_{2}^{+}$  state, respectively. Note that the statistics are computed based on the beam intensity, required beam time, and the calculated cross sections. A 10\% uncertainty (7\% systematic uncertainty and 7\% statistic uncertainty) is assumed for the deduced cross sections from the gamma spectrum. The expected accuracy for the deduced excitation energy of the 0$_{2}^{+}$ state is better than 200 keV.

%% file: STRASSE_tracker.tex
\subsection{Overview}

The STRASSE tracker consists of 6 modules and is placed inside a reaction chamber as shown in Fig.~\ref{fig:strasse_overview}(a). Each STRASSE module includes an inner and an outer layer of DSSDs as presented in Fig.~\ref{fig:strasse_overview}(b). Custom low-mass and low-capacitance microcables are used to transmit the signal from the silicon strips to the ASIC chips, called STS-\textsc{Xyter}\,\cite{xyter_nim}, which are wire-bonded to the front-end PCBs (FEBs). For STRASSE, there will be in total 4 different types of FEBs,
containing 2 (FEB2), 3 (FEB3) or 11 (FEB11$_{i}$ and FEB11$_{o}$) STS-\textsc{Xyter} chips. FEB2 (FEB11$_{i}$) and FEB3 (FEB11$_{o}$) are used to read out signals from the parallel (perpendicular) strips of the inner and outer layer respectively. Six cooling blades are used to support and cool the FEBs. A cooling base, in which the liquid coolant will circulate, is in direct contact with the cooling blades to dissipate the heat load generated by the electronics. Two star-shaped stainless steel support are used to maintain the silicon detectors. The main flange supports the whole system and includes feedthroughs for cooling pipes and connectors. It features a central hole with a diameter of 155 mm to allow the insertion of the LH$_{2}$ target.

The conceptual design of the STRASSE system has been converged based on the above-mentioned considerations and simulations. In this section, after a brief overview of the technical concept of the full STRASSE system, we present the current technical solutions and design choices considered for each of its components: detectors, electronics, cables, mechanics, integration at the RIBF and cooling.

\begin{figure}[htpb!]
\centering
\includegraphics[width=0.49\textwidth]{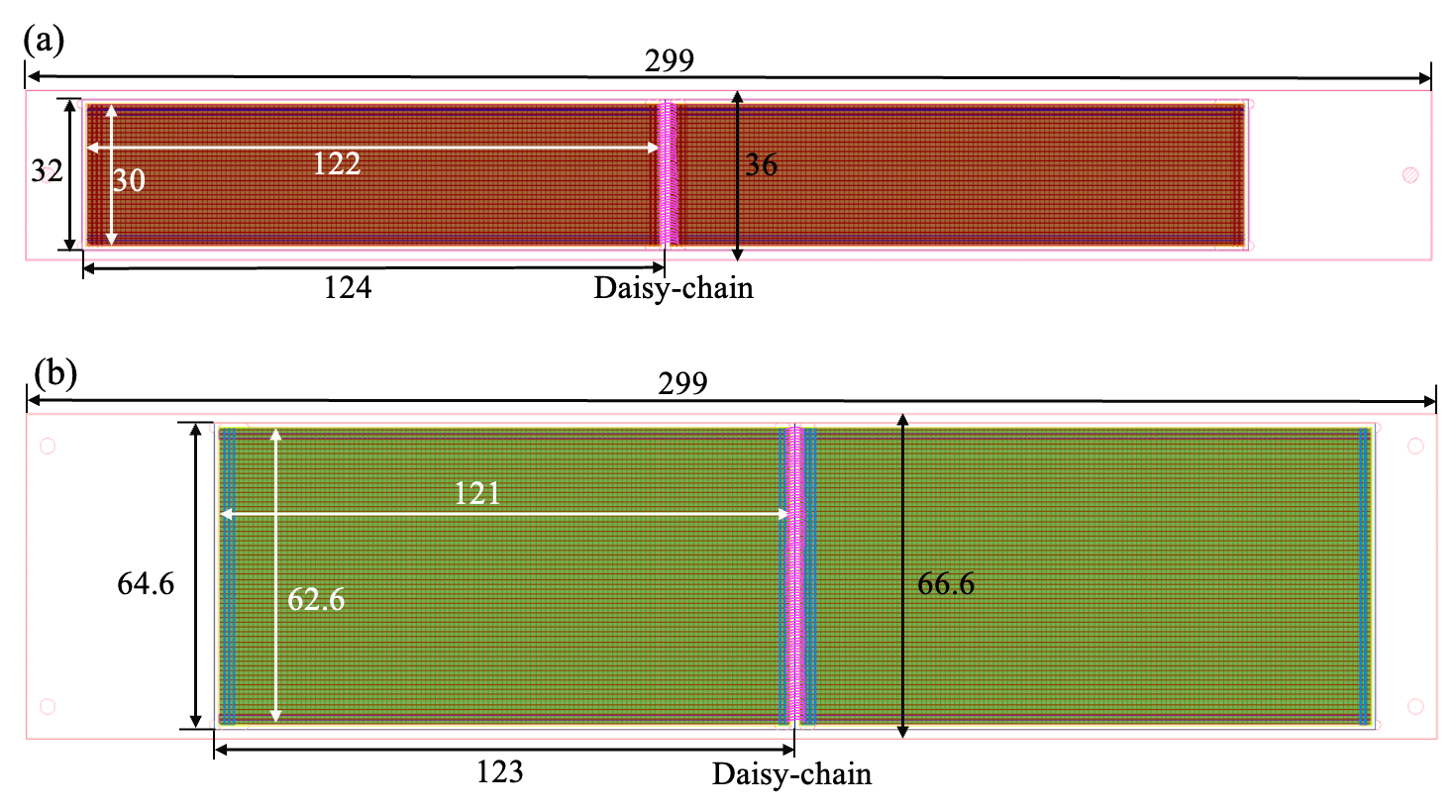}
\caption{Schematic geometry of (a) inner and (b) outer silicon sensors and their PCB frames.}
\label{fig:si_sensor}
\end{figure}

\subsection{Silicon sensors}

The array of silicon sensors consists of two detection layers called inner and outer barrel. For each layer, six detector segments are configured in a hexagonal shape (see Fig.~\ref{fig:mechanics_concept} for details). Each detector segment is made of a PCB frame holding two DSSDs as displayed on Fig.~\ref{fig:si_sensor}. One side of each DSSD corresponds to strips perpendicular to the beam axis (shorter strips) and the other side to strips parallel to the beam axis. These longitudinal strips are daisy chained between two adjacent sensors to allow their readout on a single end. A single inner silicon sensor has an active area of 30 mm $\times$ 122 mm with a thickness of 200\,$\upmu$m, while an outer silicon sensor has an active area of 62.6 mm $\times$ 121 mm with a thickness of 300\,$\upmu$m. They have the same pitch size of 200\,$\upmu$m, leading to 17358 electronic channels of the whole silicon tracker. An inactive silicon area of 1 mm surrounds the active part of the sensor.

Concerning the PCB frame, a compromise had to be made to minimize the width of the long ledges on which the silicon detectors sit while maintaining sufficient mechanical rigidity. Indeed lateral PCB ledges create dead areas in between segments of the hexagon [$\phi$ angle losses as shown in Fig.~\ref{efficicency_interaction} (right)]. After mechanical tests of a few dummy PCB frames and exchanges with the manufacturer, we decided to use 1.9\,mm-wide ledges on both sides of the sensors and 2.4\,mm-thick PCB frames made of Rogers laminate material, and not standard FR4, due to its higher mechanical rigidity.

To minimize noise, AC-coupled detectors that block leakage current from the amplifier are chosen. Polysilicon resistors within the range of 10 to 20\,MOhm are required to achieve a relatively slow recharging time of the silicon sensor with a small voltage drop. Since the noise level of the system is proportional to the input capacitance, the total capacitance of each strip including the cables is required to be smaller than 50\,pF to reach the targeted equivalent noise charge (ENC) of 10\,keV as will be discussed in Section\,\ref{sec:electronics}. The daisy-chained strips of the inner layer have the largest capacitance given their larger length and smaller thickness compared to the outer layer. Within this design, their capacitance is estimated to be 31.2\,pF with a bulk capacitance of 21.6 pF and an inter-strip capacitance of 9.6\,pF at a strip separation of 100\,$\upmu$m. The bias voltage of the silicon sensor will be within the range 50--100 V.

\subsection{Microcables}

\begin{figure*}[htpb!]
\begin{center}
\includegraphics[width=0.95\textwidth]{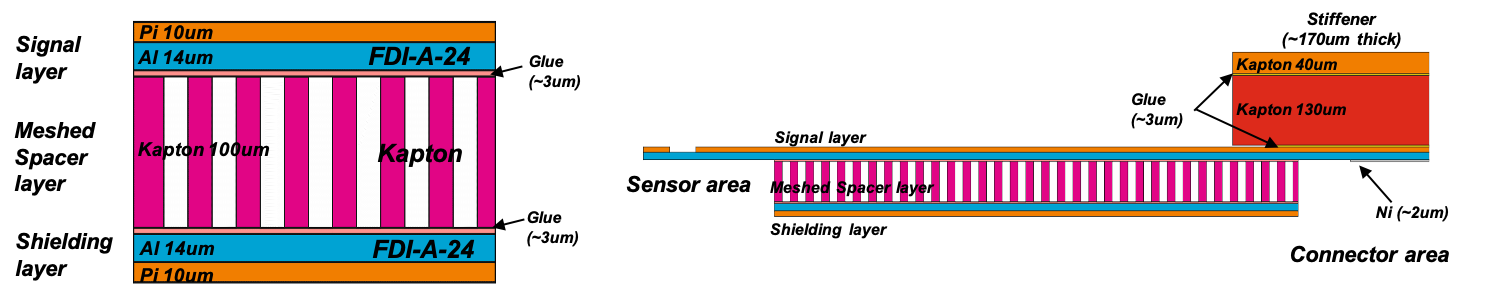}
\caption{\label{microcable_feature} Cross-section features of the LTU-STRASSE microcables. Schematic cross section of the microcable (Left) and the chain with the connection to the silicon and the connector (Right).}
\end{center}
\end{figure*}

\begin{figure}[htpb!]
\begin{center}
\includegraphics[width=0.49\textwidth]{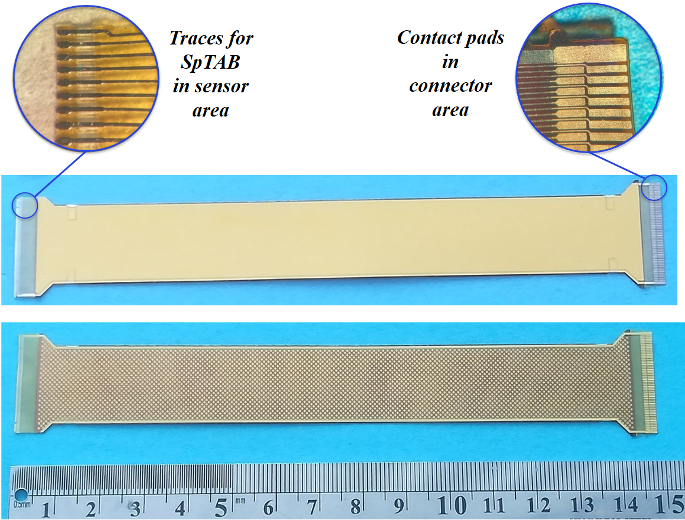}
\caption{\label{microcable_prototype} Manufactured and investigated prototype of LTU-STRASSE microcable (without connector stiffener). View of microcable from bottom layer with magnified contact areas (Top) and from top layer (Bottom).}
\end{center}
\end{figure}

\begin{figure}[htpb!]
\begin{center}
\includegraphics[width=0.49\textwidth]{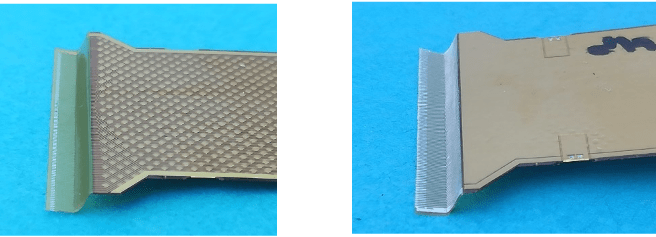}
\caption{\label{microcable_details} Folded sensor areas of microcable prototype for different interconnection options. Folded area with signal layer on top (Left) and with shielding layer on top (Right).}
\end{center}
\end{figure}

To connect the silicon strips to their front-end electronics, we will use ultra-light low-capacitance multistrip multilayered flat flexible microcables for STRASSE. These microcables are specially developed by LLL Research and Production Enterprise ``LTU" (RPE LTU) Kharkiv, Ukraine. They are made of aluminium-polyimide adhesiveless dielectrics and Kapton films. Given the complexity of the assembly (see Fig.~\ref{fig:mechanics_concept}), 15 different custom designs of microcable are necessary. All of them have a similar structure but different lengths (up to about 15 cm) and will be formed in different shapes. 

On the sensor side, the microcables will be connected by Single-point TAB (SpTAB) technique based on ultrasonic wedge welding. The combination of aluminium traces in microcables and aluminium contact pads on the sensors allow to ensure reliable monometallic joints. For mechanical robustness, SpTAB joints will be encapsulated by glue. Using aluminium conductive layers allows to realize low material budget interconnection microcables and their flexibility allows to realize volumetric arrangement of components in STRASSE. On the front-end board side, the microcables will be connected to the board using 200 $\upmu$m pitch, 50- and 120-pin HIROSE connectors (FH29 series).

Multilayered microcables consist of a main interconnection part and a connector stiffener (for matching recommended 200\,$\upmu$m thickness in contact area to the connector). Main interconnection part of the microcable includes three layers (Fig.~\ref{microcable_feature}): top signal layer, middle spacer layer and bottom shielding layer. Middle spacer layer realized as meshed one for ensuring required low-capacitance level. All layers are manufactured based on precise photolithography and chemical wet etching processes. Flexible layers are laminated together with epoxy glue to realize the required structure. The total thickness of the microcable in the main interconnection part is about 150\,$\upmu$m. Each microcable has up to 112 signal traces, the remaining traces (from 120 total possible) are used for interconnection ground and bias lines. The pitch of signal traces in the microcable is within the range 140--200\,$\upmu$m (depending on the part of the microcable) and their width is 35\,$\upmu$m.

Several microcable prototypes (Fig.~\ref{microcable_prototype}) have been developed and manufactured to investigate different design options, assembly approaches, composition and technologies for creating the final LTU-STRASSE microcables. For example, given the fact that the sensors are mounted directly on the carriers, few connection options on the sensors were studied. As visible in Fig.~\ref{microcable_details}, the prototypes were used to check that one side of the microcable can be folded in the sensor area for a reliable SpTAB connection. Importantly, the interstrip capacitance for the 15-cm long microcable prototypes are within the range of 2.2-3.7~pF (i.e. 0.15--0.25~pF/cm) matching the requirements for STRASSE.\\
 
\subsection{Electronics}

\label{sec:electronics}

\begin{figure*}[htpb!]
\centering
\includegraphics[width=\textwidth]{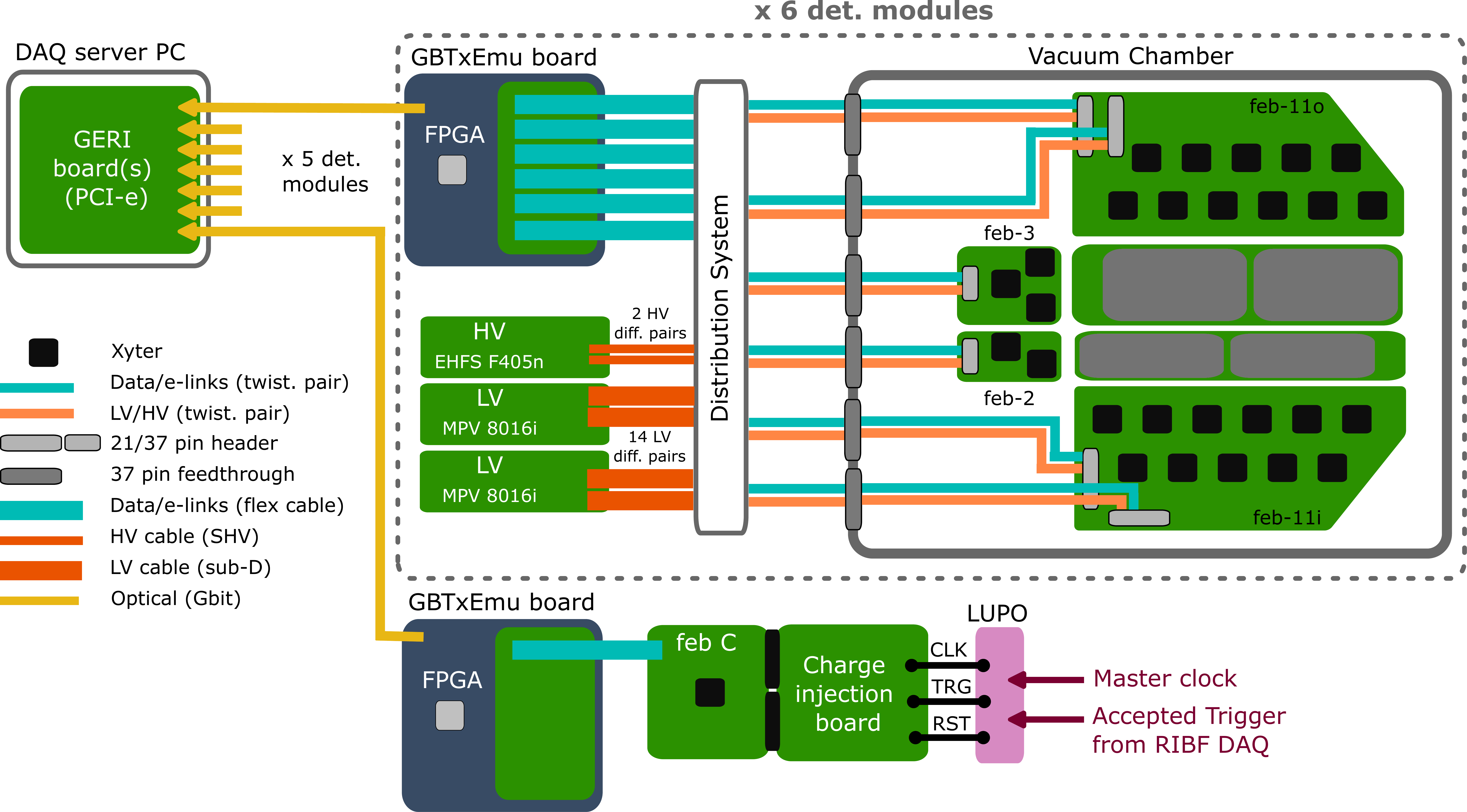} 
\caption{Simplified overview of the readout system of STRASSE. Microcables from the sensors to the front-end boards are omitted for clarity.}
\label{fig:strasse_readout}
\end{figure*}

\subsubsection{Overview}

Given that thin silicon sensors (200--300\,$\upmu$m) are necessary in STRASSE to achieve the targeted missing-mass resolution, a rather small energy deposit of protons (see Fig.~\ref{fig:penergyloss}) needs to be handled by a low-noise electronics readout chain. More specifically, the targeted ENC for STRASSE is 10 keV to keep a signal-to-noise ratio of 10 for minimal proton energy deposits of 100 keV in the array. To do so, the front-end electronics of STRASSE has to be placed as close to the silicon sensors as possible inside the vacuum chamber. As shown in Fig.~\ref{fig:strasse_overview}, the front-end boards handling the ASIC chips are placed both upstream (FEB3 and 2) and radially within the dead angles between silicon sensors (FEB11o/i) to reduce the acceptance loss. The essential requirements considered for the ASIC choice are:
\begin{itemize}
\item Compacity and low power consumption to read out the 17358 electronics channels of the STRASSE system in the vacuum environment;
\item  A dead time $\tau <$ 100\,$\upmu$s corresponding to the approximate dead time of the SAMURAI acquisition system handling other detectors of the spectrometer (limited by the readout of multi-wire drift chambers);
\item A time resolution better than 10\,ns to correlate events but a moderate energy resolution.
\item A dynamic range of 2--132 fC, to detect the protons corresponding to an energy deposit of 50 keV to 3 MeV based on simulations shown in Section\,\ref{sec:penergyloss}.
\item Demonstrated low-noise performances and successful application in nuclear physics.
\end{itemize}

The STS-\textsc{Xyter} (X,Y, Time and Energy Resolution) ASIC chip\,\cite{xyter_nim} developed by the CBM collaboration~\cite{CBMweb} suits all of the above-listed requirements and was chosen for STRASSE. The full STRASSE readout system has been adapted from the one used by CBM and is depicted schematically in Fig.~\ref{fig:strasse_readout} before entering in the details of each building block in the following sections. The STS-\textsc{Xyter} chips are placed on the front-end boards (FEB). Low-voltage differential signals (LVDS) data from STS-\textsc{Xyter} are transmitted outside of the chamber using twisted-pair cables within a custom wiring harness (under design) and 6 compact 37-pin micro-D feedthroughs for each STRASSE detector module.  Power supply (low and high voltages) will be provided via other pairs of wires within the same custom wiring harness in order to minimize the number of connectors on the FEB on which very limited space was available. Outside of the chamber, all the LVDS data from the different FEBs of a given detector module are distributed to a GBTxEMU\,\cite{lehnert2017,zabolotny2021} concentration board used for data aggregation, slow control and synchronization of the ASICs. The GBTxEMU board receives the digitized data via a mezzanine connection board (FMC) and outputs the aggregated data via optical fiber to the data acquisition server. On this server, a PCI-express GERI board\,\cite{geri} is used to catch the data frames sent by the different GBTxEMU needed for the full array and perform their synchronization. Since this readout system is triggerless, the event correlation with other detectors used in the experiment is performed by recording the timestamp of the accepted trigger of the RIBF DAQ. For that purpose, a custom charge injection board (CIB) was developed to build and transfer a timestamp from the accepted RIBF trigger signal to the STRASSE data stream via a dedicated FEB-C board with only one STS-\textsc{Xyter} chip. Details of this specific readout coupling will be presented in section\,\ref{sec:daq}.

\subsubsection{Front-end: STS-\textsc{Xyter}}

\begin{table}[htpb!]
    \centering
        \caption{Key requirements for the STRASSE readout and features of the STS-\textsc{Xyter} chip.}
    \label{tab:xyter_feature}
    \begin{tabular}{p{2.0cm}<{\centering}|p{2.5cm}<{\centering}|p{2.5cm}<{\centering}}
    \hline
    \hline
     & Requirement & STS-\textsc{Xyter} features\\
    \hline
    Number of Channel & 17358 & 128ch/chip \\
    \hline
    Polarity & Negative, Positive & Negative, Positive \\  
    \hline 
    Dynamic range & 2--132\,fC & 0--15\,fC (High-gain mode)  \\    
    & & 0--100\,fC (Low-gain mode) \\    
    \hline 
    Power consumption & Usable in vacuum & 8 mW/ch \\   
        \hline 
    Dead time & $<$100\,$\upmu$s & 0.8 $\upmu$s\\   
        \hline   
    Time resolution & $<$10 ns & 3.125 ns\\   
    
        \hline  
        \hline
    \end{tabular}
\end{table}

\begin{figure*}[htpb!]
    \centering
    \includegraphics[width=0.6\textwidth,trim={0 0 0.5cm 0},clip]{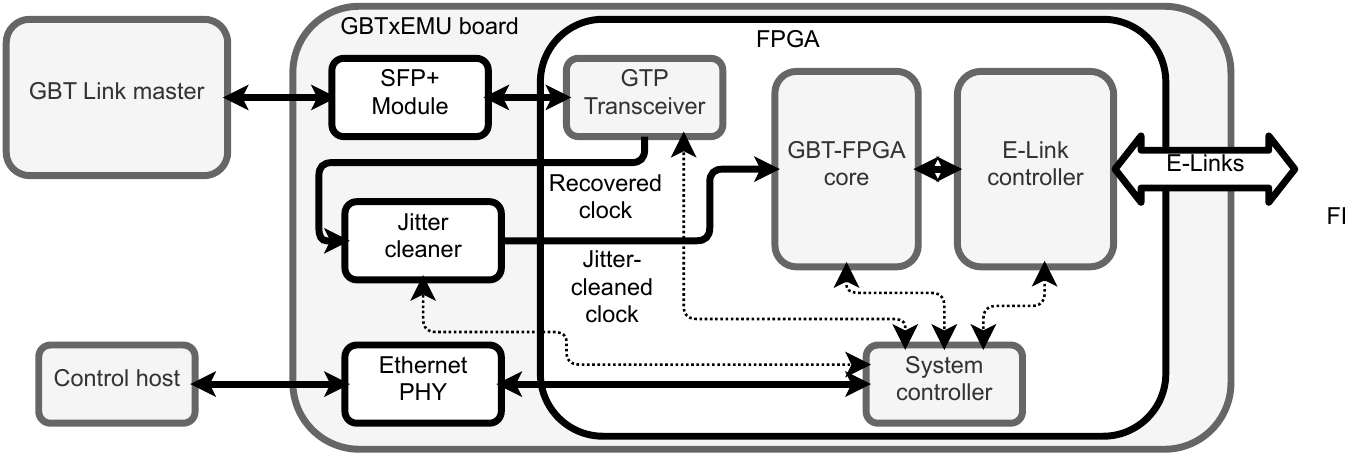}
    \includegraphics[width=0.35\textwidth, trim={0 0 20cm 0},clip]{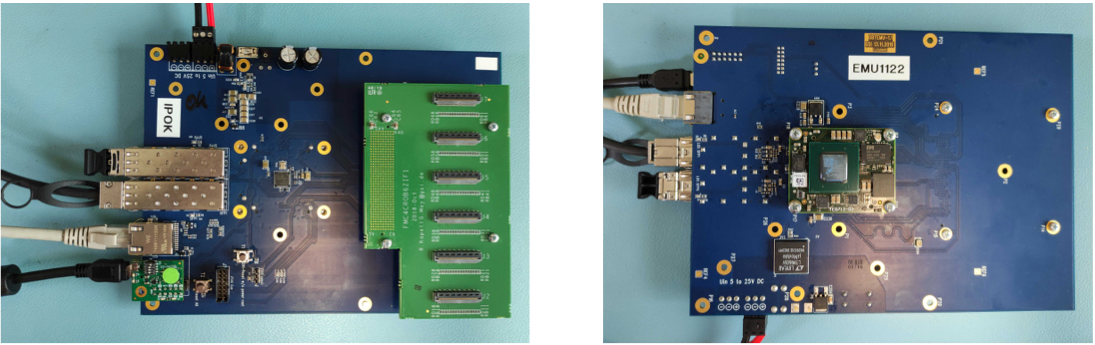}
    \caption{(Left) Simplified scheme of the GBTxEMU board from Ref. \cite{zabolotny2021}. (Right) The GBTxEMU board used for the STRASSE test setup viewed from the top side with mounted FMC board and zero-insertion force connectors on which uplinks and downlinks LVDS cables will be plugged in. A TE0712-2 FPGA module holding a Xilinx Artix 7 chip is at the center of the board on the back side.}
    \label{gbtxemu}
\end{figure*}

The main specifications of the STS-\textsc{Xyter} ASIC chip are listed in Table \ref{tab:xyter_feature} and a simplified scheme of the analog front-end of the chip can be found in Ref.\,\cite{xyter_nim}. The STS-\textsc{Xyter} chip is very compact with 128 channels in a dimension of 10~mm~$\times$~6.75 mm. The whole STRASSE system requires in total 162 STS-\textsc{Xyter} chips. They will run at a \(160\,\text{MHz}\) core clock frequency. The analog front-end part can accept positive and negative signals (corresponding to holes and electrons for a silicon detector). The system is triggerless, and a charge-sensitive amplifier (CSA) will first amplify the signal. The following polarity-selection circuit will invert the signal, in case holes are being measured. Afterward the signal is split into a fast and a slow path. For the fast path, a single-stage CR-RC shaper with a short peaking time is used to determine the signal arrival time using a leading edge discriminator. During data taking, a busy flag will be raised to be notified of potentially missed events. The 14-bit timestamp is then transferred to the back-end electronics. In the slow path, a filter of two-amplifiers with a CR-(RC)$^2$ characteristics is used for low noise energy determination and measurement. The signals pass a series of 31 resistors and are assigned a \(5\) bit ADC value. After each resistor, a threshold comparator determines if the ADC value is increased by one, until the comparison returns negative. 
Each of the comparators can be calibrated using variable resistors and an internal pulser, which is able to generate pulses in a charge of 0--100\,\text{fC}. 

To operate STS-\textsc{Xyter} properly, a 1.2 V low-voltage supply is required for the analog circuits and 1.8 V is for the digital parts. Low-voltage dropouts (LDOs) placed on the FEBs are used to regulate the input voltage from 2.4 V to 1.8 V and 1.2 V with an accuracy of 1\% and to remove noise.

\subsubsection{Back-end: GBTxEMU and GERI}

For data aggregation, a concentration board called GBTxEMU\,\cite{zabolotny2021} is used to send time-deterministic commands to STS-\textsc{Xyter} on a downlink via the LVDS type connection, as well as to receive command responses and data from STS-\textsc{Xyter} via uplink LVDS. Each STS-\textsc{Xyter} has two uplinks, while one downlink can control up to 8 STS-\textsc{Xyter} chips. The GBTxEMU board allows to emulate a real GBTX ASIC\,\cite{Leitao2015}, developed for radiation-hard environments such as LHC or CBM, with cheaper and more widely available components for less demanding experimental conditions such as STS-\textsc{Xyter} offline tests. A conceptual scheme and a picture of the board are shown in Fig.~\ref{gbtxemu}. More details on the exact references of the commercial baseboard used, the commercial FPGA used to perform the emulation and clock treatment can be found in Ref\,\cite{zabolotny2021}. To send and receive all the LVDS data signals from/to all the 4 FEBs of a detector module, a mezzanine card with all the necessary connectors is plugged in the VITA57-FMC port of the GBTxEMU board. An example of the mezzanine card considered for STRASSE is shown in green on the right part of Fig.~\ref{gbtxemu} with 6 zero-insertion force (ZIF) connectors, sufficient to readout 4 FEBs. In total, the full STRASSE silicon tracker requires 6 GBTxEMU boards and 1 additional GBTxEMU board is needed to record the trigger from the RIBF DAQ (details in Section 3.5.1).  

Finally, all the GBTxEmu boards output their aggregated data to the backend GERI concentration board hosted on the DAQ server via optical links with 4.8 Gb/s transmission speed (called GBT links). The GBTxEMU cards transmit 24-bit frames continuously on the downlinks and uplinks using 8b/10b encoding. Individual hits are sent in separate frames. The detailed frame types and formats are defined in Ref.\,\cite{GBTxManual}.  The GERI board is also used to synchronize different GBTxEMU cards.\

\subsubsection{Electronics validation}

\begin{figure*}\centering
\includegraphics[height=0.29\textwidth]{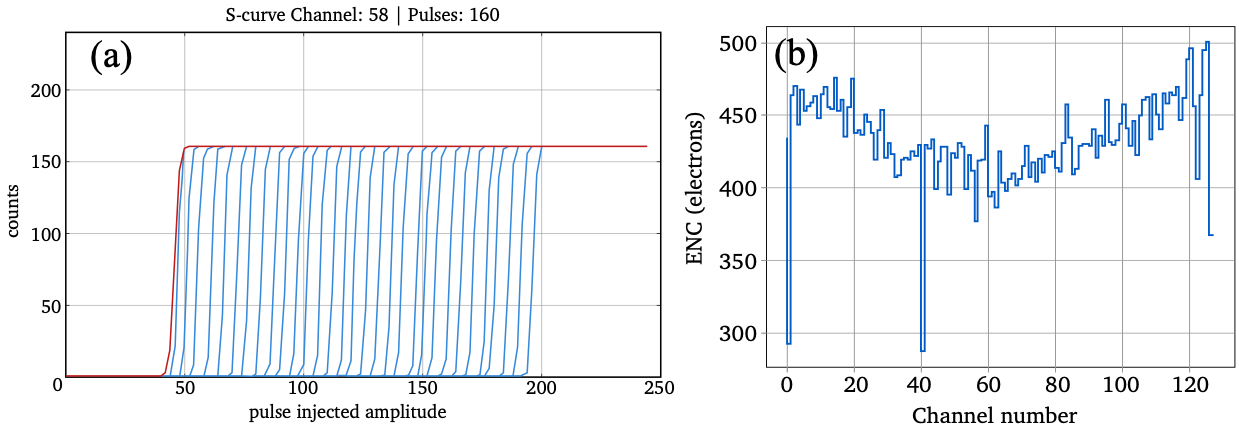} 
\caption{Overview of the S-Curve of Channel 126 (a) and the ENC spectrum (b) of the \textsc{Xyter} chip with uplink 0.}
\label{fig:scurvenoiseFeb-6}
\end{figure*}

STRASSE aims to be able to detect particles at the MIP level. The lower energy loss of 125 MeV protons in 200 $\upmu$m Si sensor to consider is $\sim$100\,keV. Assuming the mean-ionisation energy of 3.6\,eV in silicon, the signal would generate $\sim$ 28000 electron-hole pairs. To properly detect the signal, we require a signal-to-noise ratio of 10 and consequently an ENC of $\sim$\,2800 e$^{-}$ for the full readout chain including the sensor. Given the ENC performances published by the CBM collaboration, it is theoretically possible to reach this total ENC value assuming that: (i) the contribution of the STS-\textsc{Xyter} on a FEB used in vacuum can be kept below $\sim$\,500 e$^{-}$, (ii) the slope of the ENC as a function the input capacitance remains around 27 e$^{-}$/pF and (iii) the capacitance of the longest sensor strips and their microcables remains below 50\,pF.
We briefly present in this section the first tests performed at TU Darmstadt to validate the first two hypothesis (i) and (ii) and start benchmarking the STRASSE readout concept in vacuum without the final sensors.

The simple test setup consisted essentially of a STS-\textsc{Xyter} chip on a FEB-C test board located inside a vacuum chamber and mounted on a copper cooling block to dissipate the heat. On this FEB-C board a custom injector PCB was plugged in to allow connection to different input capacitors or a one-channel silicon diode. A custom-made flange hosting two PCBs was made to transmit the power supply for the STS-\textsc{Xyter} on the FEB (+2.4\,V) and the signals from the FEB in vacuum to an AFCK (AMC FMC Carrier Kintex) readout board\,\cite{afck} in air via a 40-pin twisted-pair ribbon cable. More technical details on the test bench not directly relevant here, such as the digital backend electronics used, can be found in Ref.\,\cite{AxelPhd}.

Starting without any input capacitance, our goal was to determine the noise in each channel of the STS-\textsc{Xyter} chip. For this purpose,  an internal pulser was utilized to inject different charges with a range from 0 to 255, corresponding to 0--15.32 fC, which is the total dynamic range in high-gain mode. Each charge was injected 160 times into all channels of the STS-\textsc{Xyter} chip. For each channel, all 31 ADC discriminators were inspected. The ideal case without any noise would yield a step function, meaning the discriminator does not trigger below its threshold but triggers all of the 160 pulses sent, if they are above their threshold. However, in reality, there are additional noises on the baseline, and the step function would be smeared out into what is called an S-curve. The width of this curve is directly proportional to the noise seen on the baseline (assuming a perfect signal generation from the internal pulser). In addition,  all 31 discriminators expected the same noise, making the measurement reliable. An example S-Curve of Channel 126 is shown in Fig.~\ref{fig:scurvenoiseFeb-6}(a). The ENC for each channel was extracted from the 5 lowest ADC discriminators, leading to a maximum ENC of 500\,e$^{-}$ for the outer channels and a decreasing ENC of 400 e$^{-}$ for the central channels, as shown in Fig.~\ref{fig:scurvenoiseFeb-6}(b). Adding an input capacitance of 22\,pF, the ENC slope was found to be stable at 24(1)\,e$^{-}$/pF. These results are consistent with the tests performed by the CBM collaboration~\cite{xyter_nim} (slope of 27\,e$^{-}$/pF with an incident of 539\,e$^{-}$ for the high gain mode) and allowed to validate an essential part of the readout chain for STRASSE.  

\begin{figure*}\centering
\includegraphics[width=0.8\textwidth]{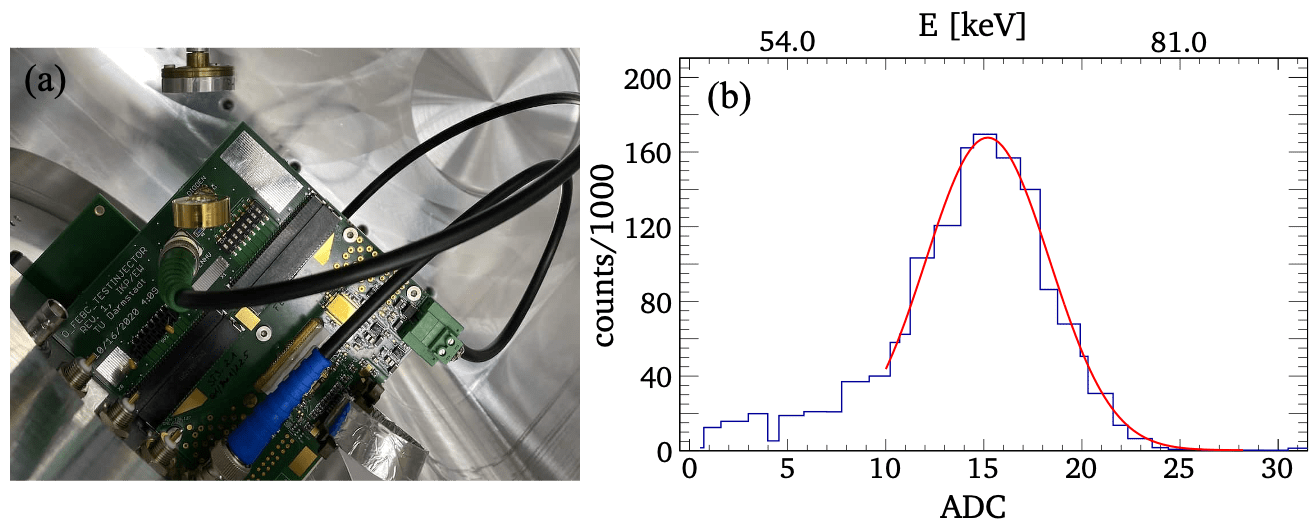} 

\caption{(a) Setup inside the vacuum chamber. The $^{241}$Am source (top, gray) was mounted on a separate stand, close to the Si junction (middle, gold). The junction was screwed onto a PCB, which was plugged into the FEB-C board. (b) The measured $\gamma$-ray spectrum using the $^{241}$Am source for a 2-hour duration with a gaussian fit (red).}
\label{fig:am_spectr}
\end{figure*}

To demonstrate the performance of the STS-\textsc{Xyter} chip and the ability to keep the ENC under control also with a Si junction connected, a measurement with an $^{241}$Am source was also performed. The $\alpha$-decay of $^{241}$Am  yielded an excited state of $^{237}$Np, whose de-excitation $\gamma$-ray of 59.54 keV was measured using a silicon detector. The source had an activity of 300 kBq. In this test two Si-Junctions were used with an active area of 50 mm${}^{2}$ and a minimum sensitive depth of 300 $\upmu$m (from Ortec, type Ametek TR-015-050-300). The operating bias voltage was -120(-160)\,V for the first (second) junction. The reverse current was certified to be 0.2(0.1) $\mu$A. The measured $\gamma$-ray spectrum with the $^{241}$Am source for a 2-hour duration is shown in Fig.~\ref{fig:am_spectr}(b). The performed gaussian fit yielded a mean at an ADC channel of 15.12 corresponding to 67.7\,keV with a $\sigma$ of 3.06 corresponding to 4.13\,keV. The significant offset (14$\%$) of the measured energy might stem from the calibration done with the internal pulser, since the pulser itself was not calibrated. The measured peak width is comparable to other measurements of $^{241}$Am using a Si diode with a $\sigma$ of 3.6\,keV\,\cite{Naumov2006}. Given that the STS-\textsc{Xyter} chips will not be used to measure precise energy deposits but to track particles, these limited spectroscopic performances are sufficient.

\subsection{Data Acquisition (DAQ)}

\label{sec:daq}

\subsubsection{Integration of the STRASSE readout into RIBF DAQ}

\begin{figure}\centering
\includegraphics[width = 0.49\textwidth]{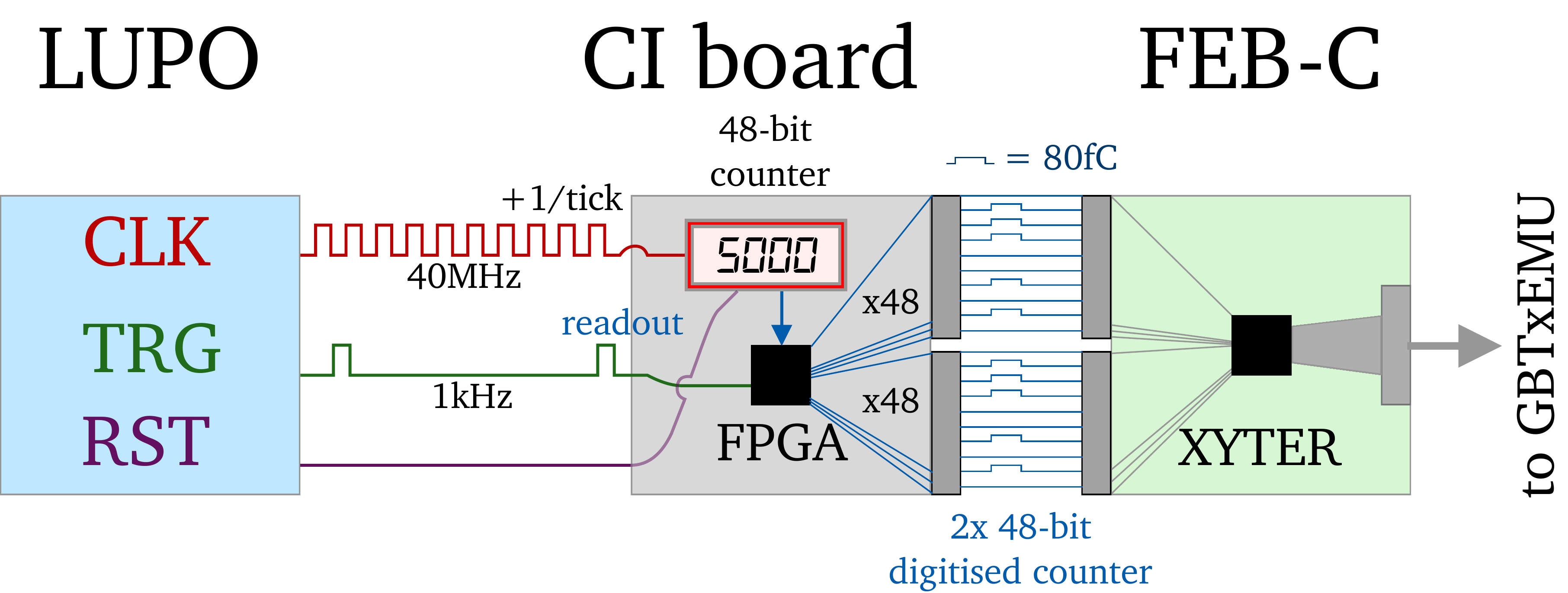}
\caption{Concept of the integration of the RIKEN DAQ into STRASSE.}
\label{fig:ciintegration}
\end{figure}

The STRASSE electronics is triggerless, i.e. an event is generated whenever a signal on a given channel crosses the threshold. The STRASSE DAQ will run in a stand-alone mode but will record the timestamp of any accepted trigger from the RIBF DAQ dealing with the other detectors.  This will allow to merge offline the different data sets by applying a coincidence window using this timestamp. This approach also allows for different coincidence time windows, to make sure no events are erroneously discarded as noise. The concept of the method developed to build, synchronize and transfer this timestamp generated by the RIBF DAQ to the STRASSE data stream via a charge injection (CI) board is shown in Fig.~\ref{fig:ciintegration}.  

First of all, a client LUPO module\,\cite{LUPO} with a dedicated FPGA add-on is used to send the accepted trigger signal of the RIBF DAQ and to transfer the master clock (CLK), generated by a separate master LUPO module (not represented in Fig.~\ref{fig:ciintegration}), to the CI board. Each time a clock (CLK) signal from the LUPO is received on the CI board, a 48-bit counter increments by one. By doing so, this 48-bit counter is equivalent to a timestamp synchronized with the RIBF DAQ.  Once the CI board receives an accepted trigger (TRG) signal, it serializes the 48-bit counter redundantly to 2 $\times$ 48 channels within 10 ns using a programmable FPGA. A logical 0 bit in the counter is represented by no output signal, and a logical 1 bit is a signal with charge Q = 80 fC, which has a length of $\sim\,1\,\upmu$s. The individual analog channels of a STS-\textsc{Xyter} chip on FEB-C receive the corresponding signals from the CI board and thus re-digitize the sent timestamp counter. Additionally, the counter can be reset if a reset (RST) signal is received from the LUPO for example at the startup of the RIBF-DAQ.

The working principle of this method has been verified by the in-beam validation experiment at HIMAC with the STRASSE demonstrator, and the details of the measurement will be presented in another paper.

\subsubsection{Trigger rates}

\begin{figure}[htpb!]
\begin{center}
\includegraphics[trim=-1cm -1.5cm 0cm 0cm, clip,width=6cm]{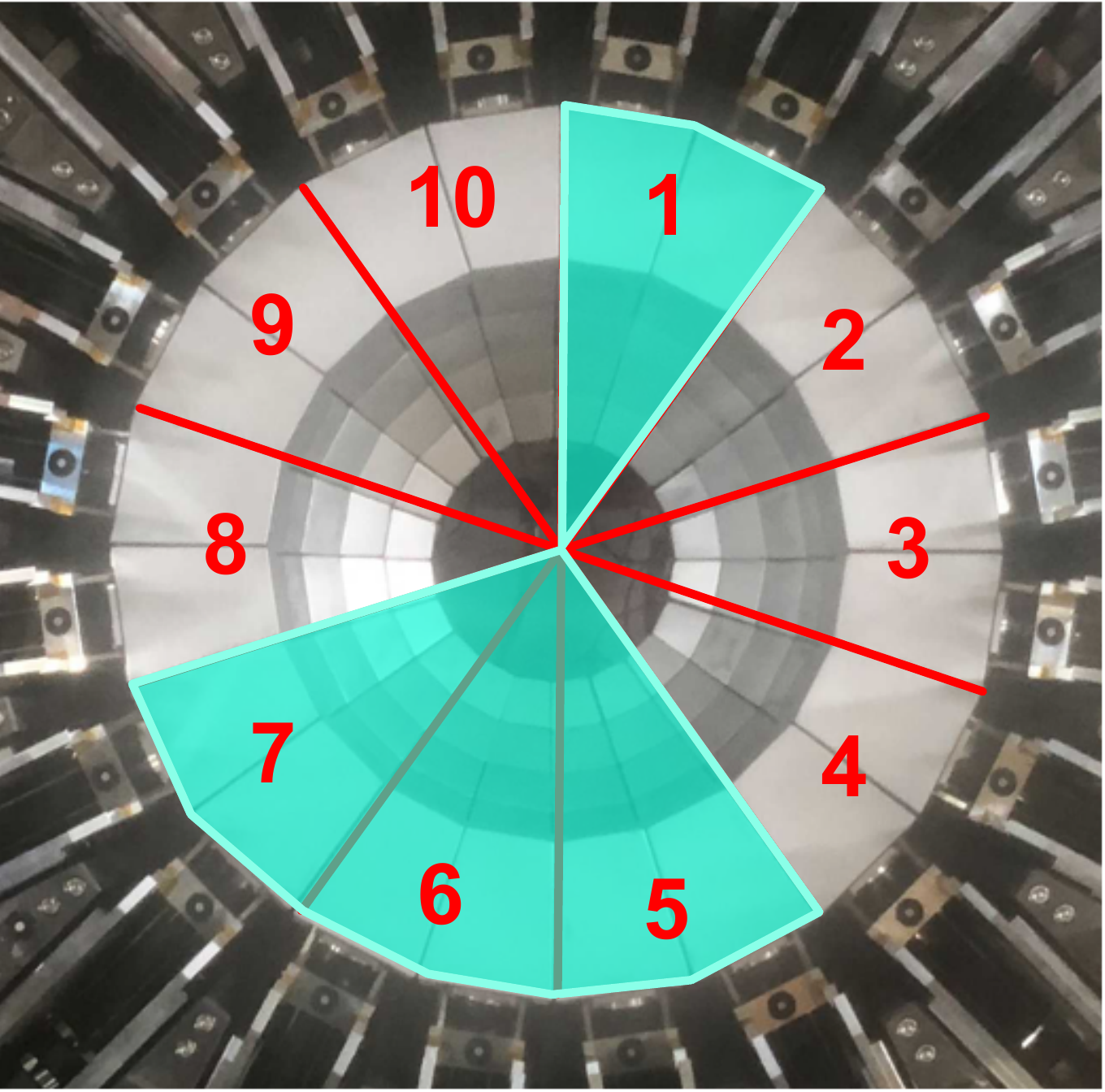}
\caption{\label{fig:trigger} Illustration of the main-trigger based on multiplicity $>1$ events with selected angular correlation in CATANA.}.
\end{center}
\end{figure}

For the missing-mass measurement, the trigger will rely exclusively on the two recoil-proton detection in CATANA, in coincidence with the beam detection in the plastic scintillator. The $(p,2p)$ reaction event in CATANA will be selected by an FPGA based trigger module, using a kinematic correlation of two protons. The two recoil protons from the $(p,2p)$ reaction have back-to-back correlation in azimuthal direction as shown in Fig.~\ref{p2pkine}, since the reaction kinematics is dominated by proton-proton elastic scattering. The event with two proton azimuthal angle $\phi = 180 \pm 36$~degrees will be selected by a new FPGA-based trigger module. This selection guarantees $>$ 99\% of the $(p,2p)$ events. A downscaled trigger with no selection on the angular correlations between the two protons will allow to monitor the rejected events. For the SAMURAI data acquisition system, the trigger rate is limited by the standard SAMURAI detectors, in particular the multi-wire drift chambers. A maximum limit of 4--5 kHz can be achieved with the newly purchased electronics and DAQ modules at SAMURAI. 

Assuming a total beam intensity of 10$^5$ pps, an inclusive $(p,2p)$ cross section of 50 mb \cite{Paul2019,Audirac2013}, a 50\% two-proton detection efficiency, a 50\% reaction loss when using a 150 mm thick LH$_{2}$ target, one gets a trigger rate of 0.65 kHz. A more reliable estimation is obtained from a full Geant4 simulation of $(p,2p)$ and INCL-based fragmentation events in the STRASSE target with a realistic CATANA geometry and efficiency, as well as the foreseen trigger with detection thresholds of 30 MeV for the individual CATANA crystals. The total counting rates in CATANA is estimated to be 1.9 kHz, among which 1.2 kHz are expected to come from $(p,pn)$ and will be fully removed by the requirement of the detection of the two tracks (correlated with the triggering CATANA crystals) in STRASSE. The resulting trigger rate is thus 0.7 kHz, including 0.62 kHz $(p,2p)$ events, well below the trigger rate limit of 4--5 kHz.

\subsection{Mechanical design}

\begin{figure*}[htpb!]
\includegraphics[width=0.95\textwidth]{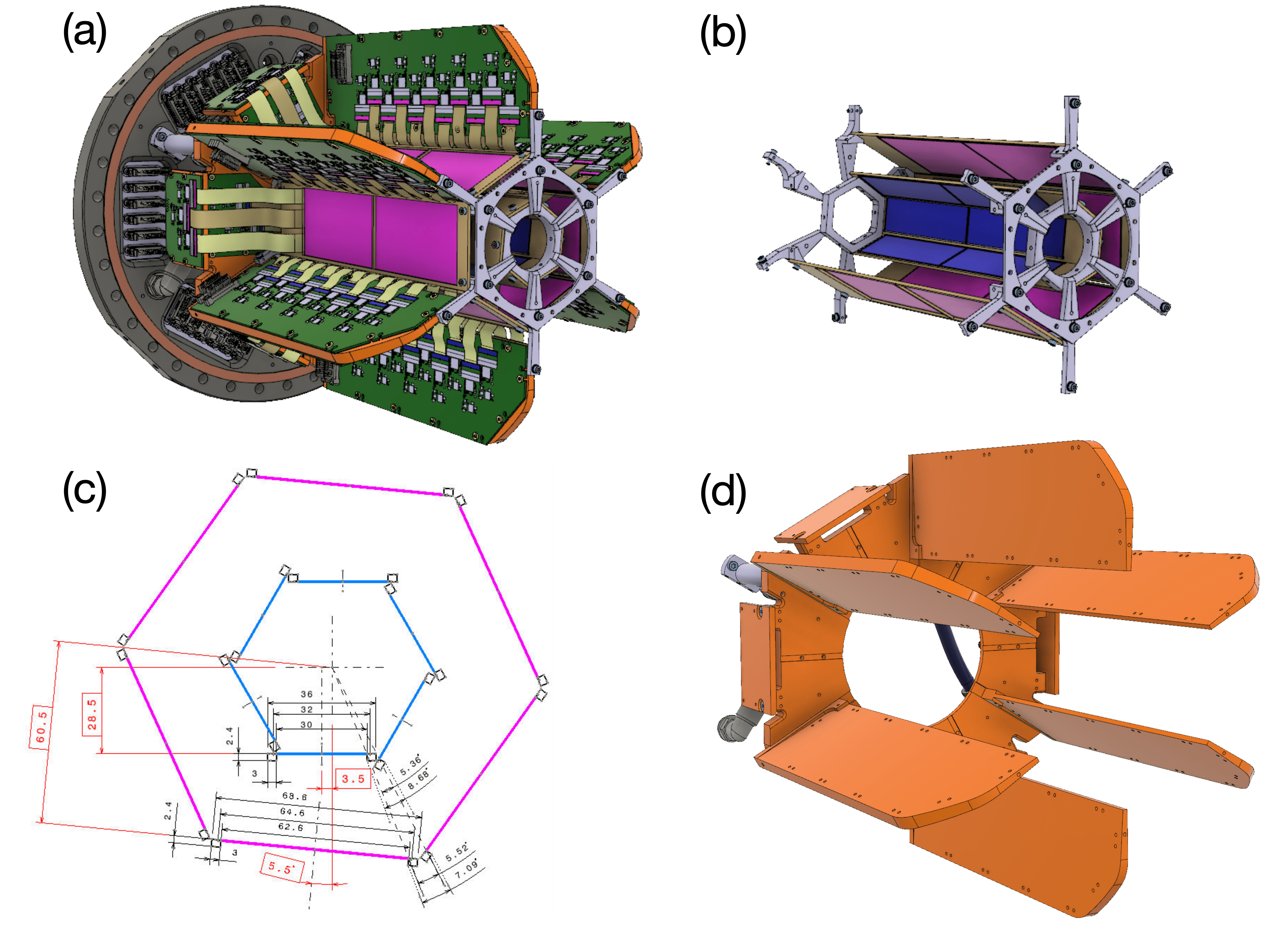}
\caption{Overall representation of the conceptual mechanical design for STRASSE. (a)\,Full array assembly.(b) Inner and outer silicon barrels assembly with their star-shaped mechanical supports on both ends (a few silcion sensors are hidden for display purposes). (c) Details of the hexagonal arrangment between neighbouring sensors optimized to minimize dead areas due to their PCB frames. (d) Front view of the cooling elements, including a base in contact with the coolant circuit and six blades in direct contact to hold some of the front-end electronic boards.}
\label{fig:mechanics_concept}
\end{figure*}

In terms of mechanical integration, the main challenges of the STRASSE project are related to the need to fit together the highly-segmented silicon detector array, the LH$_{2}$ target, the front-end electronics for 17358 channels, the cooling system and all the necessary cables in a cylinder of about 40 cm diameter and 60 cm length (size available within CATANA). Like most high-energy charged-particle trackers, this integration has to be achieved while minimizing dead detection areas.\\

 Based on these specifications, the overall mechanical design for STRASSE is displayed in Fig.~\ref{fig:mechanics_concept}. The full system is composed of several subparts:
 \begin{itemize}
  \item A \textbf{main flange} supporting all the other elements and including feedthroughs for cooling pipes, and connectors. It has a central hole of 155 mm diameter to allow the insertion of the LH$_{2}$ target.
  \item A \textbf{cooling base} (in orange) in contact with cooling circuit. FEBs for the readout of parallel strips will be directly mounted and cooled on this base.
  \item Six cooling \textbf{blades} will hold and cool the FEBs for the readout of perpendicular strips. These blades are in direct contact with the cooling base to dissipate the heat load. The cooling of the STS-\textsc{Xyter} ASIC chips is done by the backplane of the PCB using thermal vias in contact with the blades using a thin layer of thermal pad in between.
  \item \textbf{Two star-shaped stainless steel pieces} at each extreme of the array (upstream and downstream) used to hold all the silicon detectors of inner (blue) and outer (pink) barrels as shown in Fig.~\ref{fig:mechanics_concept}(b). These structural pieces are attached to both ends of the blades and will be 3D-printed. Detector PCBs will be screwed directly on them.
  \item A cylindrical \textbf{reaction chamber}, not displayed in Fig.~\ref{fig:mechanics_concept} but discussed in the next section.
  \end{itemize}
In such a design, the full array can be artificially decomposed into six STRASSE modules.

To further optimize the proton detection efficiency, the inner and outer barrel geometries and their relative positioning were optimized so that the dead angles due to the PCB frame and inactive edges of the silicon areas are minimal. This led to the configuration displayed on Fig.~\ref{fig:mechanics_concept}(c) in which:
\begin{itemize}
    \item two adjacent inner detectors were slightly offset so that neighboring detector frame edges are kept within the same dead inactive angle 
    \item the outer barrel is rotated of 5.5 degrees with respect to the inner barrel to keep PCB frames in the shadow defined by the inner dead areas.
\end{itemize} 
This configuration of the detector array originates from an efficiency optimization procedure based on multiple simulations with the \emph{nptool}~\cite{nptool} package in which we included a completely modular parametrization of the detector frame geometry (edges, ledge, thickness, offset between silicon sensors, etc). This specific configuration led to the geometrical efficiency of $49\%$ for the detection of two protons, mentioned in Section~\ref{sec:efficiency} when using the $^{17}$F$(p,2p)^{16}$O reaction at 250\,MeV/nucleon and a realistic beam profile.

Given the fixed geometry of the silicon detector array discussed above and the radius of the reaction chamber, the FEBs and cooling blades for perpendicular strips place radially in these dead angles. Their dimensions have been maximized to fit in all the STS-\textsc{Xyter} ASICs, passive components and connectors for the microcables but also for cables going to the flange (HV, LV, and digital signals). For the perpendicular strips the FEBs have a surface of approximately $9 \times 30$ cm, and $9 \times 7$ cm for the parallel strips.

\subsubsection{Reaction chamber}

\begin{figure*}[htpb!]
\includegraphics[width=0.95\textwidth]{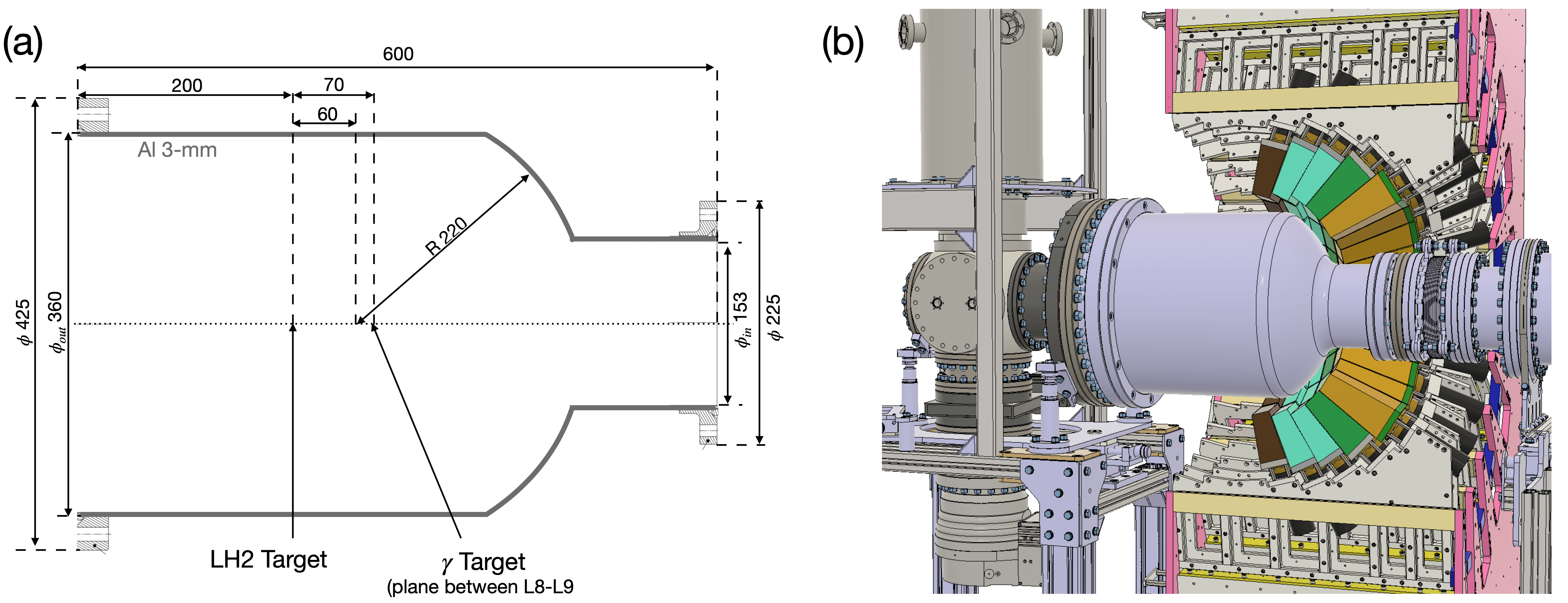}
\caption{Reaction Chamber: (a) dimensions and target positions (b) 3D view within CATANA (not all crystal layers are represented) and assembled with the cryogenic target on the left.}
\label{fig:mechanics_chamber}
\end{figure*} 

The reaction chamber geometry was mainly designed to fit within the CATANA array but also keeping in mind, later, the possible use of STRASSE in conjunction with high-resolution gamma-ray spectrometers for spectroscopic measurements. As a result, the dimensions obtained [detailed in Fig.~\ref{fig:mechanics_chamber}(a)] allows to maximize the space available to fit the array and the FEBs in a cylinder of 360 mm while keeping a distance between the center of the LH$_{2}$ target to the chamber walls ranging from 180 mm at 90$^{\circ}$ to about 260 mm at forward angles.

\begin{figure*}[t!]
\centering
\includegraphics[width=0.95\textwidth]{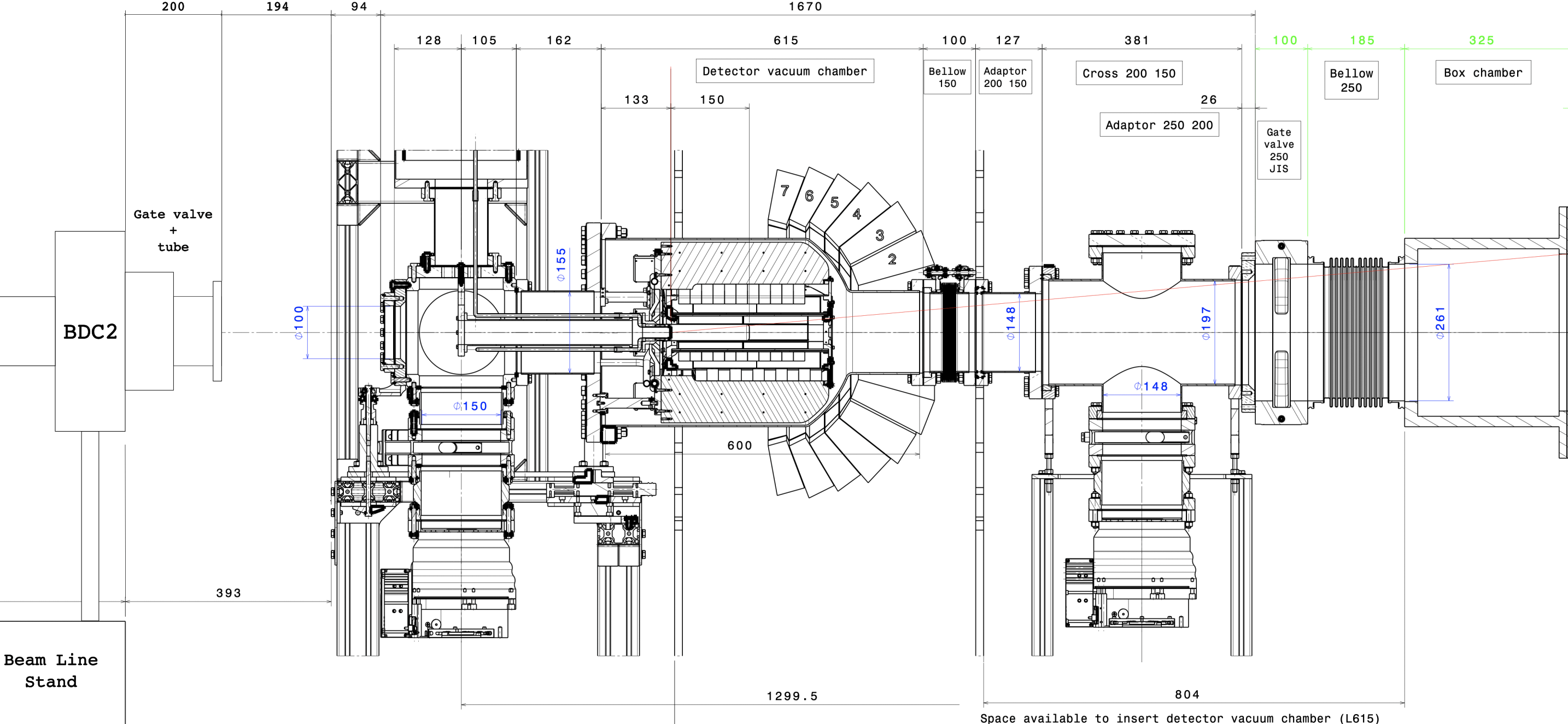}
\caption{Secondary target area of the SAMURAI spectrometer with a preliminary integration of the full STRASSE+CATANA system.}
\label{fig:samurai}
\end{figure*}

The chamber is composed of a main body (large cylinder + curved surface + small cylinder) and two flanges support (entrance and exit). The main constraint is to be as thin as possible while staying rigid enough to sustain the vacuum constraint. Several materials and manufacturing options for the main body of the chamber have been studied. Deformation calculations under vacuum were performed and indicate that a uniform 3-mm thickness aluminium chamber of the above-mentioned shape can sustain the mechanical stress. Thickness uniformity depends on the manufacturing process, as a consequence, vacuum test and thickness measurements will be performed with a first version of the chamber built by cold forming (metal spinning) a 6 mm thick Al sheet. If these tests are unsatisfactory, full machining of a single aluminium block is a possible alternative but inevitably more costly.

\subsection{Integration on the SAMURAI beamline at the RIBF}

\subsubsection{SAMURAI target area}

STRASSE will be mainly used together with the CATANA array at the SAMURAI target area. SAMURAI is a large acceptance spectrometer \cite{SAMURAI1,SAMURAI2} based on a superconducting dipole magnet at the RIBF. It has a momentum resolution of $\emph{p}$/$\sigma_p \sim 700$, leading to a clean ($\geq 3 \sigma$) mass separation up to mass $A\sim 100$. The typical SAMURAI setup is composed of high-efficiency and high-resolution beam and fragment trackers upstream and downstream the secondary target, high-efficiency NEBULA neutron detector array and large-area hodoscopes to trigger on charged particles transmitted to the focal plane of the magnet as detailed in Ref.\,\cite{SAMURAI1}. 

The secondary target location of SAMURAI is a space of about 1.9\,m  between the beam and fragment tracking devices (BDCs and FDCs, respectively)  which can be arranged to the needs of the experiments. In Fig.~\ref{fig:samurai} we present a preliminary implantation of the STRASSE and CATANA systems at this location. From left (upstream) to right (downstream), we fit the cryogenic target frame and its entrance chamber ($\simeq$\,35\,cm), STRASSE+CATANA ($\simeq$\,1\,m) and a second pumping station ($\simeq$\,40\,cm). The cryogenic taget will be installed first with the structure supporting the empty flange of STRASSE and the silicon array will be inserted from downstream on a temporary translation system with CATANA open. Finally, the second pumping station, sitting on a rail system perpendicular to the beam axis, will be inserted and coupled to the rest of the beamline.  Note that the STRASSE system will be isolated with windows from the beamline pipes.\\

\subsubsection{CATANA configuration}
CATANA is a CsI(Na) scintillator array aimed originally to measure $\gamma$ rays from exotic nuclei but also suitable for recoil proton measurement. It was updated in 2020 by adding 40 crystals. The CsI(Na) crystals are arranged into 7 rings (L2-L8, in total 140 crystals) and cover polar angles from 17 to 77 degrees. One crystal covers 7$\sim$ 9 degrees of polar angle and 18 degrees of azimuthal angle. CATANA can measure the kinetic energy of protons up to 250 MeV. Each crystal is read by an independent PMT (HAMAMATSU R580). In order to detect both the $\gamma$ rays from reaction residues and the recoil protons from the QFS reactions, a dual gain readout of CATANA has been developed. The base circuits of PMTs have to be modified to take the signals from the anode and the second last dynode of PMTs. The anode signal is amplified and digitized by ADC/TDC for the $\gamma$-ray measurement, while the dynode signal is directly read by a pulse digitizer for proton measurement. Note that the CATANA crystals lying at L2 will be operated in single-gain mode and used to detect only protons. For the 662 keV $\gamma$ ray from the $^{137}$Cs source, the energy resolution and the photo-peak efficiency are measured to be 9\% in FWHM and 21.5\% after addback (17.6\% before addback), respectively. \\

\subsection{Cooling}
\label{sec:cooling}

\begin{figure}[t!]
\begin{center}
\includegraphics[width=0.3\textwidth]{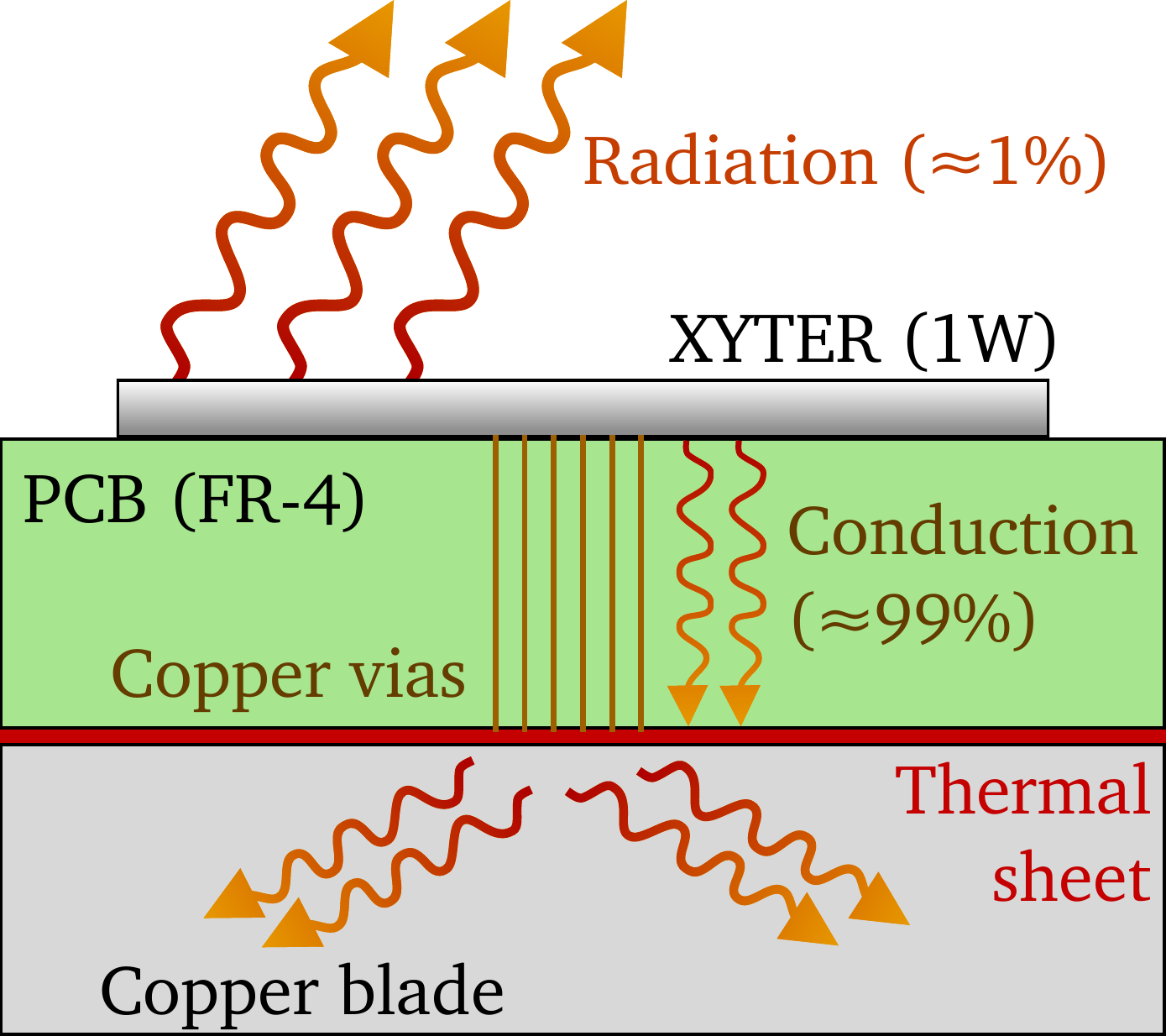}
\includegraphics[width=0.45\textwidth]{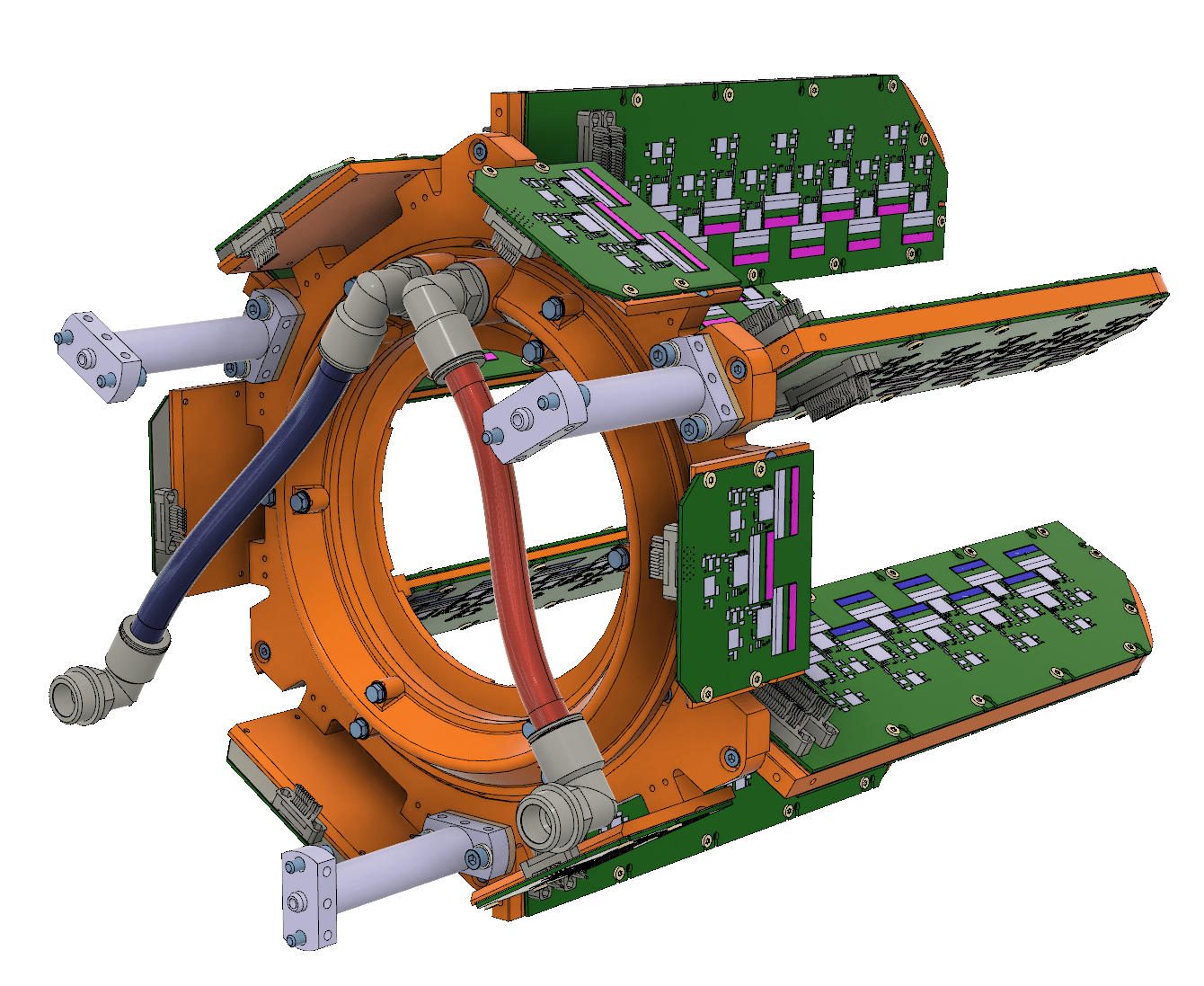}
\includegraphics[width=0.45\textwidth]{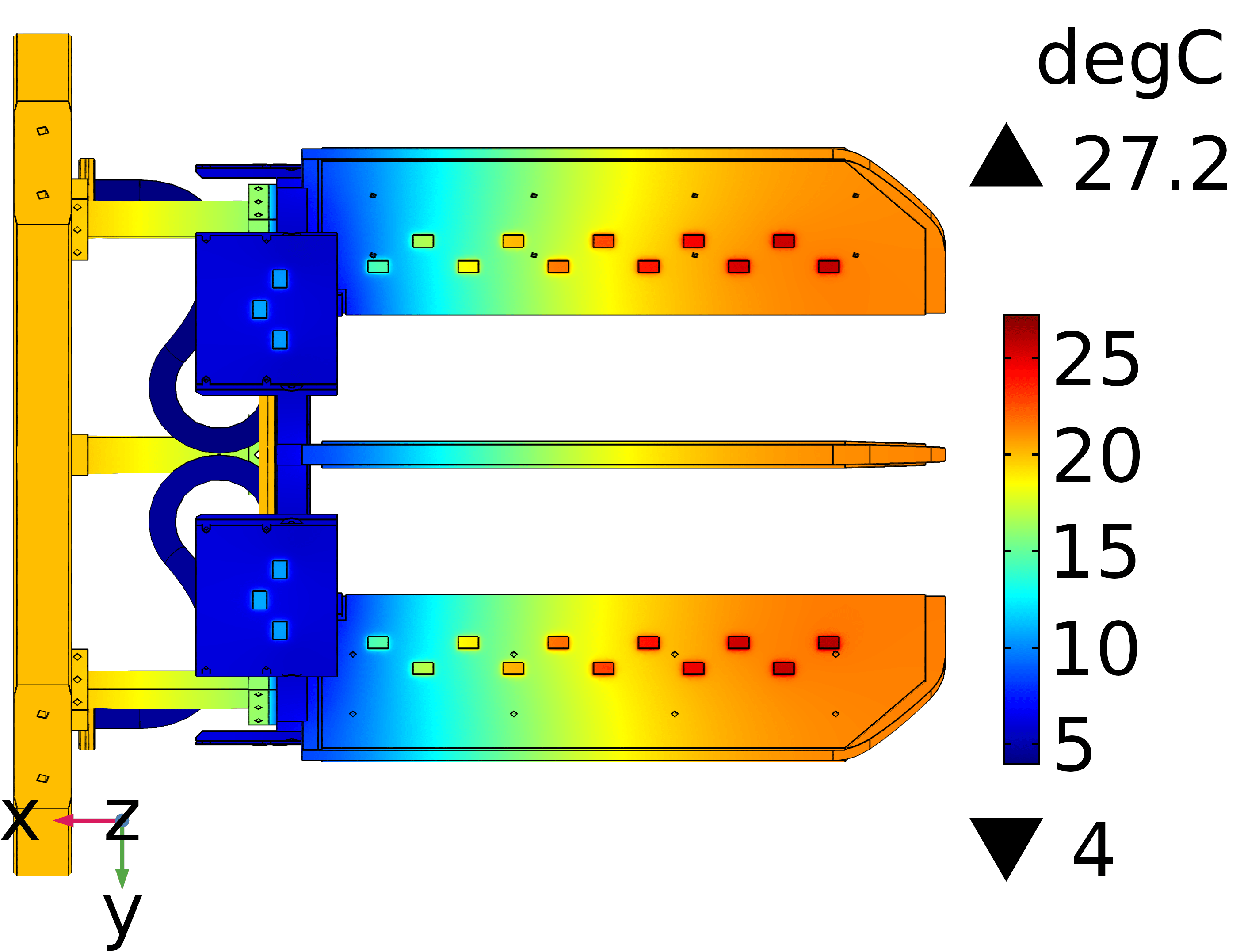}
\end{center}
\caption{(Top) Transmission of heat from the  STS-\textsc{Xyter} chip. (Middle) Back view of the cooling system for STRASSE with the front-end boards atached. (Bottom) COMSOL simulation of the temperature distribution of the STRASSE system.}
\label{fig:strassecool}
\end{figure}

The cooling system aims to evacuate the power dissipated essentially by the STS-\textsc{Xyter} ASICs on the FEBs in vacuum. Glob-top is applied to the top side of the STS-\textsc{Xyter} chips to protect the bonding wires to FEBs. At $\sim$ 60 $\celsius$, the wire bondings could be detached due to the thermal expansion of the glob-top. Therefore, the chips need to work below that temperature and can only be cooled from the back side of the PCB. As shown in Fig.~\ref{fig:strassecool} (top), the heat generated by the  STS-\textsc{Xyter} chip is conducted through the 96 copper vias with a diameter of 0.3 mm to the back of the PCB. Subsequently, a thin thermally conducting but electrically insulating sheet is utilized to improve the contact surface and conduct the heat onto the copper blade.

STRASSE has 24 FEBs which carry in total 162 chips, consuming around \(166\,\text{W}\) of power. The limited space inside the vacuum chamber, as well as the optimization of the active area for the recoil protons imposes a challenge in the design of the STRASSE cooling system. COMSOL simulations \cite{comsol} have been performed to guide the design of the cooling system. As a result, six copper cooling blades are designed to maximise the cross section of the material, while minimising solid angle covered as shown in Fig.~\ref{fig:strassecool} (middle). The thickness of the blade ranges from \(4\,\text{mm}\) on the inner part to \(10\,\text{mm}\) on the outer part, covering only \(5^{\circ}\) solid angle. At the downstream side they are round to not collide with the vacuum chamber. At the upstream side, these six blades are connected to a copper cooling base which features a pipe pressed into it and allows the coolant to flow. The  STS-\textsc{Xyter} chips, which read out the perpendicular strips are distributed over both sides of the cooling blades to allow for better conduction of heat. The heat conduction is made via the contact area of \(\sim 8\,\text{cm}^{ 2}\) between each cooling blade and the cooling base, which itself measures \(15\,\text{mm}\) in thickness. The STS-\textsc{Xyter} chips responsible for the parallel strip readout are directly connected to the cooling base via individual extensions, yielding an optimal cooling. For the transfer of heat between the cooling base and the coolant flowing through the cooling pipes, one loop has been found to be sufficient as shown in Fig.~\ref{fig:strassecool} (bottom).

\begin{figure}[h]
\includegraphics[width=0.49\textwidth]{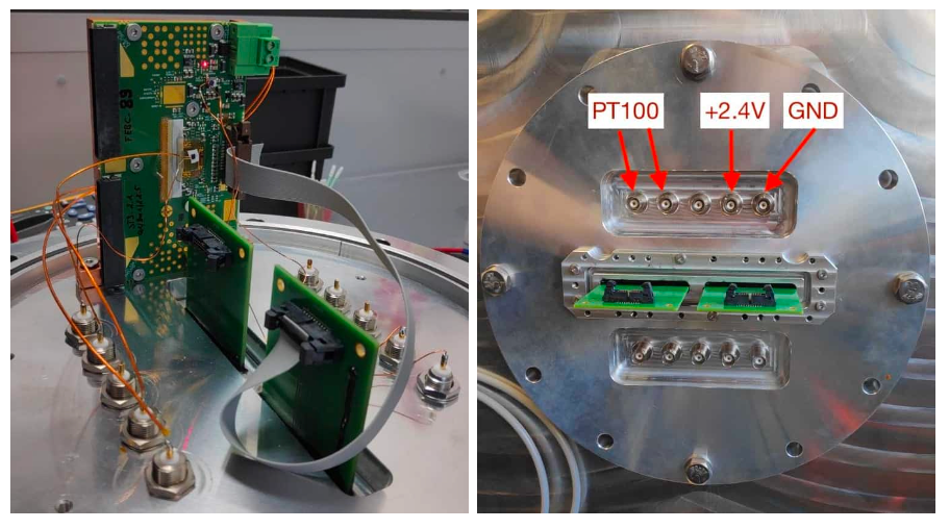}
\caption{Overview of the test setup for the thermal measurements. (Left) Inside the vacuum chamber. The setup was running, as indicated by the red lit LED. (Right) Outside of the vacuum chamber with labels.}
\label{fig:thermal_exp}
\end{figure} 

The coolant of the system was chosen to be water at a throughput of 4 L/min and an inlet temperature of 5 $^{\circ}$C. It leaves the option to fabricate the cooling blades from aluminium, limiting weight and cost, while at the same being easy to handle. The flow speed can be increased and the water can be cooled by an additional 5 K, leaving more headroom to stabilize the temperature. In Fig.~\ref{fig:strassecool} (bottom) the obtained temperature distribution for the selected parameters is shown. Note that this simulation does not contain the additional heat from the LDOs. 

In order to benchmark the COMSOL simulations, thermal measurements of one  STS-\textsc{Xyter} chip on the FEB-C card were conducted in a vacuum chamber. Fig.~\ref{fig:thermal_exp} shows the overview of the setup. The FEB-C card was fixed to a cooling block, which was then attached to the flange. In these two contact surfaces, a 0.508 mm thick thermal sheet was inserted. Two PT100 temperature sensors were attached with thermal tapes to the top of the STS-\textsc{Xyter} chip and the voltage divider region. The other side was soldered to two BNC feedthroughs. A stick temperature probe was attached to the vacuum chamber to get a reference point. All three temperature probes were then read out by a 4-channel Pico PT104. The tests were conducted in vacuum with a pressure smaller than 1.5$\times$ 10$^{-6}$ mbar. The power consumption of the FEB-C card was 1.46 W. Two different cooling blocks were used. One was made of copper, and the other was made of AlMg$_{3}$. The measured mean temperature difference of the STS-\textsc{Xyter} surface to the outside of the vacuum chamber was 6.2 K for copper and  7.8 for AlMg$_{3}$. For the COMSOL simulation, only heat transport by conduction was modeled in vacuum, since this process contributed more than 99\% of the heat transmission. To mimic any possible convection on the air side of the flange, the airflow to the flange was assumed to be 1 mm/s with an initial temperature of 21.2 $\celsius$. The simulations gave a value of 5.1 K for copper and 6.5 for AlMg$_{3}$, which were consistent with the measurement. The underestimation in $\Delta$T in the simulation likely originated from uncertainties in the material quality and the transmission of heat through the thermal vias. For the purpose to guide the design of the cooling system for the STRASSE setup, such deviation is acceptable.

%% file: LH2_target.tex
\begin{figure}[htpb!]
  \centering
  \includegraphics[width=0.45\textwidth,trim={0.1cm 0.1cm 0.1cm 0.1cm},clip]{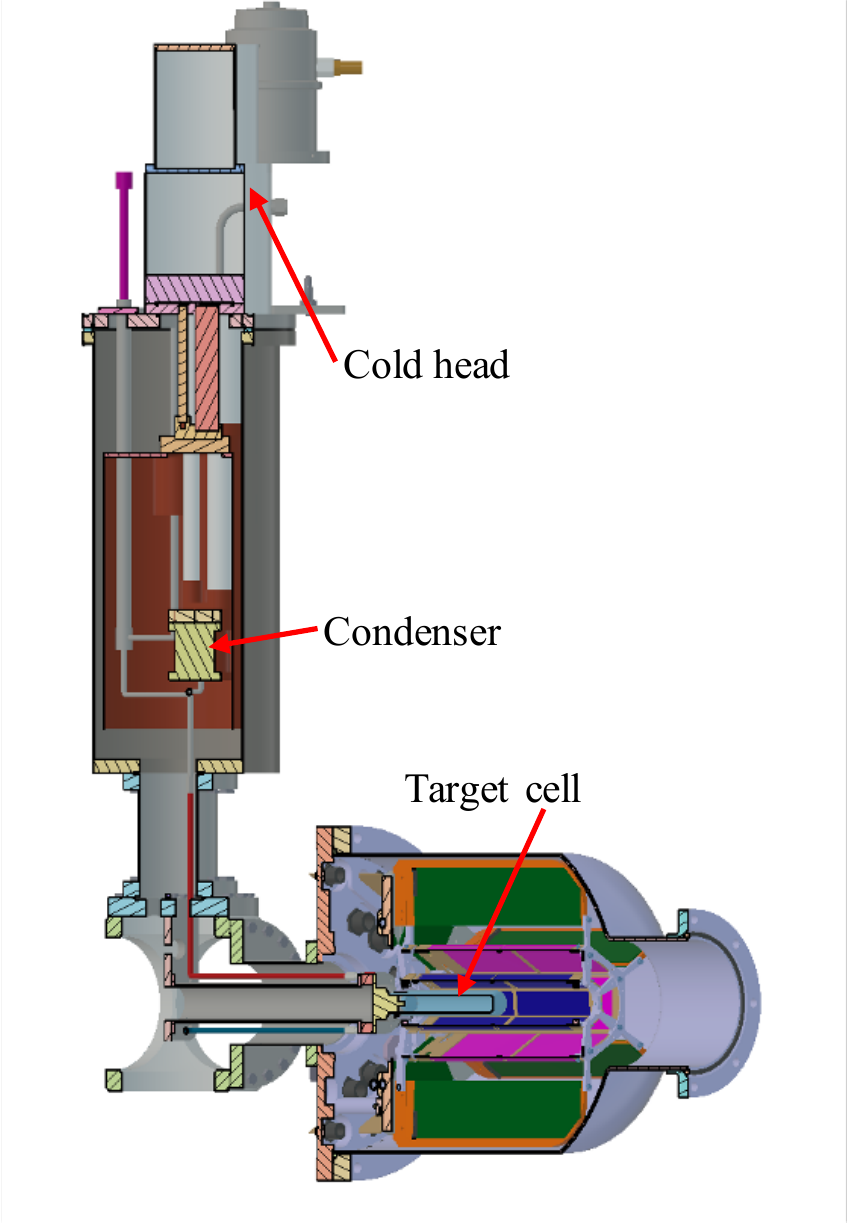}
  \caption{Cut view of the full STRASSE system including the cryostat.}
  \label{fig:Cryostat}
\end{figure}

\label{sec:lh2target}
The STRASSE tracker is meant to be combined with a LH$_2$ target with a length up to 150 mm. A pure LH$_2$ target significantly improves the luminosity for a given energy loss in the target, compared to an easier-to-use CH$_2$ target, and provides cleaner data for QFS experiments. The density of LH$_2$ depends slightly on the temperature of vapor pressure imposed, with typical values of 75 mg/cm$^{3}$ at 16 K, corresponding to $4.5\times 10^{21}$ atoms per cm$^3$. In the specific use of STRASSE, the LH$_2$ target will minimize the energy loss and angular straggling of the protons produced from quasi-free scattering, a requirement to reach an acceptable missing-mass energy resolution for a thick-target measurement. 

The cryogenic system will be composed of a compact cryostat dedicated to the liquefaction of hydrogen, the target cell itself, a storage tank for the hydrogen needed during the experiment and a control and monitoring system, following the philosophy of the former PRESPEC \cite{prespec} and MINOS targets \cite{obertelli14}. In this section, a brief overview of the cryogenic system is given, while details will be given in a forthcoming publication.

\subsection{Cryostat}
The cryostat is installed above the LH$2$ target as shown in Fig.\,\ref{fig:Cryostat}. The cold head lies at the top of the cryostat. The condenser located at about 50-cm height relative to the target cell is installed on the second stage of the cold head, which is the coldest part of the cryogenic system. The STRASSE target system is based on the thermosiphon principle.
The hydrogen gas is cooled and liquefied by contact with the condenser. Liquid hydrogen flows by gravity to the target cell, and then becomes vapors due to the heat load around the target. The cold vapors go back to the condenser via the H$_2$ return tube and are liquefied and fall again into the supply tube. Such loop continues until the target cell is full of liquid hydrogen.

The cryostat is compact with a weight of 110 kg, a total height of 1.3\,m (20\,cm in diameter) and a length of 1.0\,m. It will be capable of liquefying hydrogen and filling the target cell in less than 12\,hours. The system is conceived to empty the target cell in less than 20\,minutes to perform the empty target measurement and restore the initial experimental conditions in a short time. The system is conceived as a closed loop, guaranteeing additional safety. More than 1 psi over-pressure of the hydrogen gas will return the gas to the storage tank through calibrated check valves.

\subsection{Target cell}
The target cells are composed of polyethylene terephthalate
(PET) film, also known as the trademark Mylar.
Their external diameter will be 31 mm with an entrance window of 20-mm diameter. This reduced diameter contributes to the optimization of the missing-mass resolution. The target length can be chosen up to 150\,mm, corresponding to $\sim$ 6.4 $\times$ 10$^{23}$~protons/cm$^{2}$. The total volume of liquid hydrogen in the system is $\sim$ 0.25\,L.

The cylinder tube is obtained by thermo-molding a rectangular sheet of Mylar with a thickness of 175\,$\upmu$m and then glued to the two sides in order to form a tubular shape. The end cap is thermo-pressed and the molded piece is glued at the end of the tube part. The entrance window is obtained in a similar manner as the end cap. The fabricated target cells have been tested under high pressure with water and they could stand a pressure of over 11 bars.

%% file: Conclusions.tex
STRASSE is a new charged-particle silicon tracker combined with a liquid hydrogen target to be used at the RIBF facility of RIKEN, in particular at the SAMURAI experimental area, for quasi-free scattering experiments at 200--250 MeV/nucleon. Its compact geometry in vacuum will combine missing-mass and prompt $\gamma$-ray spectroscopy measurements. For missing mass studies, a resolution better than 2 MeV is expected for $(p,2p)$ and $(p,3p)$ quasi-free scattering measurements with the CATANA scintillator array. The simulated detection efficiency for the detection of both protons from $(p,2p)$ is 49\%. Thanks to the vertex tracking with a position resolution better than 1 mm, the in-beam $\gamma$ spectroscopy resolution will be limited only by the $\gamma$-detector array. With a high-purity Ge tracking array, a Doppler-corrected energy resolution of 1\%--2\% could be reached. Thanks to its STS-\textsc{Xyter} based readout system initially developed for the CBM experiment, STRASSE is expected to be operational at rates up to 1 MHz. These features will open new physics opportunities for nuclear physics with radioactive beams available at the RIBF.